\documentclass[review]{elsarticle}
\usepackage{bm}
\usepackage{amsmath}
\pdfoutput=1
\usepackage{amssymb}
\usepackage{pgfplots}
\usepgfplotslibrary{fillbetween}
\usetikzlibrary{spy}
\pgfplotsset{compat=1.10}
\usepackage{amsbsy}
\usepackage{graphicx}
\usepackage{subcaption}
\usepackage{mathtools}
\usepackage[export]{adjustbox}
\usepackage{dashbox}
\usepackage{array}
\newcolumntype{L}[1]{>{\raggedright\let\newline\\\arraybackslash\hspace{0pt}}m{#1}}
\newcolumntype{C}[1]{>{\centering\let\newline\\\arraybackslash\hspace{0pt}}m{#1}}
\newcolumntype{R}[1]{>{\raggedleft\let\newline\\\arraybackslash\hspace{0pt}}m{#1}}
\DeclarePairedDelimiter\ceil{\lceil}{\rceil}

\newcommand{\xx}{\mbox{\boldmath $x$}}
\def\bn{\mbox{\boldmath $n$}}
 
\usepackage{lineno,hyperref}
\usepackage{latexsym}
\usepackage{moreverb}
\usepackage{tikz}
\usepackage{setspace}
\usetikzlibrary{decorations.markings}
\usetikzlibrary{matrix,shapes,arrows,positioning,chains,shapes.geometric,calc,patterns,
decorations.pathmorphing,decorations.markings}
\usepackage{float}
\usepackage{natbib}
 \usepackage{graphics}
\usepackage{graphicx}
\DeclareGraphicsExtensions{.eps,.pdf,.png,.jpg,.jpeg,.xcf}
\usepackage{epsfig}
\usepackage{amssymb}
\usepackage{multirow}
\usepackage{amsmath}
\usepackage{wasysym}
\usepackage{graphicx}
\usepackage{psfrag}
\usepackage{amsbsy}
\usepackage{epstopdf}
\usepackage{subcaption}
\usepackage{amsfonts}
\usepackage{setspace}
\usepackage{color}
\usepackage{float}
\usepackage{caption}
\usepackage{tabularx}
\usepackage{epsfig}
\usepackage{bm}
\usepackage{dcolumn}
\usepackage{gensymb}
\usepackage{pdfpages}
\graphicspath{{Pics/}} 
\DeclareGraphicsExtensions{.png,.jpg}

\makeatletter
\long\def\@maketablecaption#1#2{\@tablecaptionsize
    \global \@minipagefalse
    \hbox to \hsize{\parbox[t]{\hsize}{\centering #1 \\ #2}}}

\makeatother

\def\vec#1{\mbox{\boldmath $#1$}}

\usepackage[english]{babel}
\usepackage{algorithm}

\def\b1{\mbox{\boldmath $1$}}

\def\bC{\mbox{\boldmath $C$}}

\def\bG{\mbox{\boldmath $G$}}

\def\bK{\mbox{\boldmath $K$}}
\def\bM{\mbox{\boldmath $M$}}
\def\bN{\mbox{\boldmath $N$}}

\def\bR{\mbox{\boldmath $R$}}

\def\bb{\mbox{\boldmath $b$}}

\def\bn{\mbox{\boldmath $n$}}
\def\bp{\mbox{\boldmath $p$}}

\def\bu{\mbox{\boldmath $u$}}

\def\bnabla{{\mbox{\boldmath $\nabla$}}}

\def\bu{{\vec u}}

\def\Otnm1f{\Omega^\mathrm{f}(t^{\mathrm{n}-1})}
\def\Otnm12f{\Omega^\mathrm{f}(t^{\mathrm{n}-\frac{1}{2}})}

\def\Otnm1s{\Omega^\mathrm{s}_\mathrm{i}(t^{\mathrm{n}-1})}
\def\Otnm12s{\Omega^\mathrm{s}_\mathrm{i}(t^{\mathrm{n}-\frac{1}{2}})}

\def\testf{\vec \phi^\mathrm{f}}

\def\dO{\mathrm{d}{\Omega}}

\def\usnp1{{\vec u}^\mathrm{s,n+1}}

\def\vphinp1{\vec{\eta}^\mathrm{s,n+1}}

\def\Oe{\Omega^\mathrm{e}}

\def\bC{\mbox{\boldmath $C$}}

\def\bG{\mbox{\boldmath $G$}}
\def\bK{\mbox{\boldmath $K$}}
\def\bM{\mbox{\boldmath $M$}}

\def\bR{\mbox{\boldmath $R$}}

\def\Gma{\mbox{\boldmath $\Gamma$}}

\def\b1{\mbox{\boldmath $1$}}

\def\bR{\mbox{\boldmath $R$}}

\def\bn{\mbox{\boldmath $n$}}

\def\bu{\mbox{\boldmath $u$}}

\def\bnabla{{\mbox{\boldmath $\nabla$}}}

\def\bu{{\vec u}}

\def\Otnm1f{\Omega^\mathrm{f}(t^{\mathrm{n}-1})}
\def\Otnm12f{\Omega^\mathrm{f}(t^{\mathrm{n}-\frac{1}{2}})}

\def\Otnm1s{\Omega^\mathrm{s}(t^{\mathrm{n}-1})}
\def\Otnm12s{\Omega^\mathrm{s}(t^{\mathrm{n}-\frac{1}{2}})}

\def\testf{\boldsymbol{\phi^\mathrm{f}}}

\def\dO{\mathrm{d}\boldsymbol{\boldsymbol{\Omega}}}

\def\testf{{\vec \phi}^\mathrm{f}}

\newcommand{\nwc}{\newcommand}

\nwc{\qref}[1]{(\ref{#1})} 

\nwc{\ip}[1]{\langle #1 \rangle}

\nwc{\ta}{\tilde{a}}

\newcommand{\bs}{\boldsymbol{\sigma}}

\def\bu{{\vec u}}

\def\bn{\mbox{\boldmath $n$}}

\def\Otnm1f{\Omega^\mathrm{f}(t^{\mathrm{n}-1})}
\def\Otnm12f{\Omega^\mathrm{f}(t^{\mathrm{n}-\frac{1}{2}})}

\def\Otnm1s{\Omega^\mathrm{s}(t^{\mathrm{n}-1})}
\def\Otnm12s{\Omega^\mathrm{s}(t^{\mathrm{n}-\frac{1}{2}})}

\def\testf{\vec \phi^\mathrm{f}}

\journal{arXiv}
\begin{document}

\begin{frontmatter}

\title{An Efficient Deep Learning Technique for the Navier-Stokes Equations: Application to Unsteady Wake Flow Dynamics}


\author[mymainaddress]{T. P. Miyanawala}

\author[mymainaddress]{R. K. Jaiman\corref{mycorrespondingauthor}}
\cortext[mycorrespondingauthor]{Corresponding author}
\ead{mperkj@nus.edu}

\address[mymainaddress]{Department of Mechanical Engineering, National University Singapore, Singapore 119077}

\begin{abstract}
We present an efficient deep learning technique for the model reduction of the Navier-Stokes equations for unsteady flow problems. 
The proposed technique relies on the Convolutional Neural Network (CNN) and the stochastic gradient descent method.
Of particular interest is to predict the unsteady fluid forces for different bluff body shapes at low Reynolds number. 
The discrete convolution process with a nonlinear rectification is employed to approximate the mapping between the bluff-body shape and the fluid forces. 
The deep neural network is fed by the Euclidean distance function as the input and the target data generated by the full-order Navier-Stokes computations for primitive bluff body shapes. The convolutional networks are iteratively trained using the stochastic gradient descent method with the momentum term to predict the fluid force coefficients of different geometries and the results are compared with the full-order computations. 
We attempt to provide a physical analogy of the stochastic gradient method with the momentum term with the simplified form of the incompressible Navier-Stokes momentum equation. We also construct a direct relationship between the CNN-based deep learning and the Mori-Zwanzig formalism for the model reduction of a fluid dynamical system.
A systematic convergence and sensitivity study is performed to identify the effective dimensions of the deep-learned CNN process such as the convolution kernel size, the number of kernels and the convolution layers. Within the error threshold, the prediction based on our deep convolutional network has a speed-up nearly four orders of magnitude compared to the full-order results and consumes an insignificant fraction of computational resources. The proposed CNN-based approximation procedure has a profound impact on the parametric design of bluff bodies and the feedback control of separated flows.
\end{abstract}


\end{frontmatter}


\section{Introduction}
Unsteady separated flow behind a bluff body causes fluctuating drag and transverse forces to the body, which is of great significance in engineering applications.
``After over a century of effort by researchers and engineers, 
the problem of bluff body flow remains almost entirely in the empirical, 
descriptive realm of knowledge,'' as famously stated by \cite{roshko1993perspectives}.
Recent advances in computational and experimental techniques in the past decades 
have only strengthened this statement. 
Nonlinear unsteady flow separation and the dynamics of vortex formation 
in the near wake region make the bluff-body flow an extremely complex phenomenon
to tackle through rigorous analytical tools.  
While physical experimental and computational techniques provide 
high-fidelity data, they are generally time-consuming and expensive for design space exploration and flow control in a practical engineering application. 
Furthermore, the enormous amount of generated high-fidelity data are often under-utilized and the significant patterns identified and learned in one case are rarely 
used for a next simulation effectively. For example, consider the flow past 
a stationary rounded corner square cylinder and the engineering objective 
is to determine a direct relationship between the fluid forces and the rounding angle of the square cylinder.  
Traditionally, we need to perform computationally expensive transient Navier-Stokes simulations for each rounded angle value for a considerable number of time-steps, albeit the vortex formation and the shedding process do not differ much between the different configurations. 
Here, we focus on a data-driven computing method to predict the key 
flow parameters of bluff-body flows via available high-fidelity data. 
Of particular interest is to have an efficient alternative for the brute force full-order model (FOM) and to obtain the fluid forces at nearly real-time within an acceptable error threshold while utilizing a fraction of the computational effort. 

Data-driven methods using low dimensional representations via proper orthogonal decomposition (POD) \cite{holmes2012turbulence}, 
eigenvalue realization algorithm (ERA) \cite{yao2017model} and dynamic mode decomposition (DMD)\cite{schmid2010dynamic,schmid2011applications} have been used 
in computational fluid dynamics (CFD) for model order reduction. 
These methods generate physically 
interpretable spatial and spatio-temporal modes related to dominating features 
of the bluff body flow such as the shear layers and the Karman vortex street. 
However, these techniques may fail to capture nonlinear transients and multi-scale phenomena  \citep{kutz2017deep} in bluff-body flows,
which motivates the need for an improved technique to extract invariant dominant features to predict the important flow dynamical parameters. 
In such learning algorithms, the essential idea is to construct a generalized kernel mapping function from a finite set of output data and to postulate a priori relationship between the input-output dynamics.
With that regard, biologically-inspired learning (i.e., neuron-based learning) techniques have some attractive approximation properties to construct the input-output dynamics 
through two main phases namely, training and prediction.
In the training phase, high-fidelity data from FOM are fed to a learning algorithm 
as pairs of input and output. The target of neural network learning 
is then to find the functional relationship between the input (bluff body geometry)
and the output (wake dynamics) using the provided initial data such that 
the force coefficient for a new bluff body geometry can be determined in
real time without using the full-order simulation. 
For this process, the learning algorithm utilizes a set of weighting/functional 
steps also called layers to connect the input geometry matrix and the force coefficient. These hidden layers, the set of input geometry matrices and the output
force coefficient layer form a \emph{neural network}. After the input data is fed 
to the neural network, it initializes the training with guessed 
weights for the layers between the input and output and iteratively 
back-propagates the deviations and corrects the assumed 
weights. In \cite{parish2016paradigm} a modeling paradigm was proposed using field inversion and machine learning for fluid mechanics problems. However, in these traditional machine learning algorithms, neural network layers utilize matrix multiplication by considering a matrix of parameters with another parameter and each data point in one layer is connected
to each data point of the next layer via a weight function. These early 
supervised learning methods, also referred to as shallow learning,
performs reasonably well if the system is governed by a few dominant features 
and becomes extremely inefficient in capturing the multiple local 
flow features such as flow separation, shear layer, vortex formation region, and street.  

Deep learning solves the deficiency of traditional learning algorithms by building complex features from simpler nested functions via input-output relationships.  
For example, \cite{ling2016reynolds} proposed a novel Reynolds averaged turbulence model based on deep learning.
%
Another developing technique is machine learning based on Gaussian process (GP) regression whereby the interpolations are governed by previous covariances opposed to smooth curve-fitting. The GP-based regression network 
replaces the weight connections
in a Bayesian neural network with Gaussian processes, which allows the model input dependent correlations between multiple tasks.
Such GP-based models are generally task-specific and may require sophisticated approximate Bayesian inference which is much more computationally expensive in contrast to deep learning models.
Recently, the GP-based machine learning method is introduced 
for the linear differential equation systems by \cite{raissi2017machine} 
and for the nonlinear partial differential equation (PDE) systems by \cite{raissi2017hidden}. Also, 
\cite{parussini2017multi} introduced a multi-fidelity GP for prediction of random fields. 
While neural networks employ a considerable number of highly adaptive basis functions, 
Gaussian processes generally rely on a multitude of many fixed basis functions.
Another main difference of CNN-based technique with the GP-based regression method is that it allows 
a parametric learning process through highly adaptive basis functions.
Neural networks have some inherent ability to discover meaningful high-dimensional data through learning multiple layers via  highly adaptive basis functions \cite{Hinton2006deep,Bengio2009}. 
In particular, deep neural networks can provide an effective 
process for generating adaptive basis functions to develop 
meaningful kernel functions, which can allow discovering structure in high-dimensional data without human intervention.
Owing to these attractive properties of neural networks, 
we consider a CNN-based based deep learning to develop a novel computational framework to predict the wake flow dynamics using 
high-fidelity computational fluid dynamics (CFD) data. 

The present work aims to develop an efficient model reduction technique for the Navier-Stokes equations 
via  deep neural networks. 
In a deep convolutional network framework \citep{lecun1998gradient}, the 
output data point of a layer is connected only to a 
very small portion of the input data points and the learning system provides the sparse interactions (sparse weights) and parameter sharing (tied weights) for working with inputs of variable size.
Due to this local connectivity between the adjacent layers, a 
particular learned weight function can properly capture the essential features with much reduced memory requirements and computing operations. 
While the CNN applies a discrete convolution operation, it is local in extent and performs superior to other deep learning methods.
Hence, we consider the convolutional neural network 
as the deep-learning technique to learn and predict the wake flow parameters. 
Our main contribution is to develop the feedforward CNN-based prediction model and to assess its accuracy and performance against the traditional Navier-Stokes simulation for unsteady bluff-body separated flow. 
Another key contribution is to establish a relationship of deep kernel 
learning method with the simplified form of the incompressible Navier-Stokes equations, which essentially provides a physical justification for the success of the proposed neural networks during the functional mapping between the input and the output response. We also connect the convolution 
process in the deep learning procedure with the well-known Mori-Zwanzig formalism for the dimensionality reduction of a dynamical system.
%
For a demonstration of the proposed data-driven technique, 
we consider a simple variation of bluff body geometry as a parametric input 
and perform a series of experiments to demonstrate the accuracy 
and the efficiency of the technique.
The proposed novel model order reduction (MOR)-technique based on deep learning has direct  engineering applications in the design of offshore, civil and aeronautical structures.

The paper is organized as follows. We begin by reviewing the full-order model based on the variational formulation of the unsteady incompressible flow equations in Section 2. Section 3 describes the background material and the formulation of our CNN-based deep learning procedure and covers the relationship of deep-learning 
based MOR with the analytical form of the full-order Navier-Stokes system.
In Section 4, we present a systematic sensitivity and performance analysis 
of the proposed CNN-based technique in predicting the wake dynamics 
of various bluff bodies of different geometries.
Concluding remarks are presented in Section 5.

\section{Full-Order Variational Model}
For the sake of completeness, we briefly summarize the Navier-Stokes solver used in this numerical study for our high-dimensional full-order computations. We assume the fluid flow to be Newtonian, isothermal and incompressible for the present study.
A Petrov-Galerkin finite-element and semi-discrete time stepping are adopted for the full-order modeling to investigate the interaction of an incompressible viscous flow with a stationary body \citep{jaiman2016,jaiman_caf2016}. 
For the spatial domain $\Omega$ and the time domain $(0,T)$, the incompressible Navier-Stokes equations for the momentum and the continuity are 
\begin{align}
\rho\left ( \frac{\partial \bu} {\partial t} 
+ \bu\cdot \boldsymbol{\nabla} \bu \right) = \boldsymbol{\nabla} \cdot
{\boldsymbol{\sigma}} + \boldsymbol{b} 
, \qquad
\boldsymbol{\nabla} \cdot\bu = 0 \quad
\mbox{ on } \Omega \times (0,T)
\label{eq:NS}
\end{align}
where $\bu=\bu(\xx,t)$ represent the fluid velocity defined for each spatial point $\xx \in  \Omega$, respectively, $\boldsymbol{b}$ is the body force applied on the fluid, and ${\boldsymbol{\sigma}}$ is the Cauchy stress tensor for a Newtonian fluid, written as:
${\boldsymbol{\sigma}} = -{p} \boldsymbol{I} + \mu \left(\boldsymbol{\nabla} \bu + \left(\boldsymbol{\nabla} \bu\right)^T\right)$,
where ${p}$ denotes the fluid pressure, $\mu$ is the dynamic viscosity of the fluid. 
The appropriate conditions are specified along the Dirichlet $\Gamma_g$
and Neumann $\Gamma_h$ boundaries of the fluid domain.
We next present the discretization using a stabilized  variational procedure with equal order interpolations for velocity and pressure. 
 
\subsection{Petrov-Galerkin finite element for fluid flow}
By means of the finite element method, the fluid spatial domain $\Omega$ is discretized into several non-overlapping finite elements $\Oe$,
$\mathrm{e} = 1, 2, ... , n_\mathrm{el}$, where $n_\mathrm{el}$ is the total number of elements. In this paper,
we adopt a generalized-$\alpha$ method to integrate in time $t \in [t^\mathrm{n},t^\mathrm{n+1}]$, which can be unconditionally stable as well as second-order accurate for linear problems simultaneously via a single parameter termed as the spectral radius $\rho_{\infty}$. With the aid of the generalized-$\alpha$ parameters $(\alpha, \alpha_\mathrm{m})$, the expressions employed in the variational form for the flow equation are given as \cite{jansen}:
\begin{align}
	{\boldsymbol{u}}_\mathrm{h}^\mathrm{n+1} &= {\boldsymbol{u}}_\mathrm{h}^\mathrm{n} + \Delta t \frac{\partial{\boldsymbol{u}}_\mathrm{h}^\mathrm{n} }{\partial t}+ \gamma \Delta t \left( \frac{\partial{\boldsymbol{u}}_\mathrm{h}^\mathrm{n+1}}{\partial t}- \frac{\partial{\boldsymbol{u}}_\mathrm{h}^\mathrm{n}}{\partial t} \right), \\
	{\boldsymbol{u}}_\mathrm{h}^\mathrm{n+\alpha} &= {\boldsymbol{u}}_\mathrm{h}^\mathrm{n} + \alpha({\boldsymbol{u}}_\mathrm{h}^\mathrm{n+1} - {\boldsymbol{u}}_\mathrm{h}^\mathrm{n}), \\
	\frac{\partial{\boldsymbol{u}}_\mathrm{h}^\mathrm{n+\alpha_m}}{\partial t}&= \frac{\partial{\boldsymbol{u}}_\mathrm{h}^\mathrm{n}}{\partial t} + \alpha_\mathrm{m}\left( \frac{\partial{\boldsymbol{u}}_\mathrm{h}^\mathrm{n+1}}{\partial t}- \frac{\partial{\boldsymbol{u}}_\mathrm{h}^\mathrm{n}}{\partial t} \right),
\end{align}
where 
$	\alpha_\mathrm{m} = \frac{1}{2}\bigg( \frac{3-\rho_\infty}{1+\rho_\infty}\bigg), \quad
	\alpha = \frac{1}{1+\rho_\infty}, \quad
	\gamma = \frac{1}{2} + \alpha_\mathrm{m} - \alpha$.
%
Let the space of the trial solutions be denoted by $\mathcal{S}^\mathrm{h}$ and the space of test functions be $\mathcal{V}^\mathrm{h}$. The variational form of the flow equations can be written as: find $[\bu^{\mathrm{n}+\alpha}_\mathrm{h}, {p}^\mathrm{n+1}_\mathrm{h}]\in \mathcal{S}^\mathrm{h}$ such that $\forall [\testf,q] \in \mathcal{V}^\mathrm{h}$:
\begin{align}
&\int_{\Omega_\mathrm{h}}\rho\left(\frac{\partial\bu^{\mathrm{n}+\alpha_\mathrm{m}}_\mathrm{h}}{\partial t} + \bu^{\mathrm{n} + \alpha}_\mathrm{h} \cdot\bnabla\bu^{\mathrm{n}+\alpha}_\mathrm{h}\right) \cdot \testf \dO \notag
+\int_{\Omega_\mathrm{h}}\bs^{\mathrm{n}+\alpha}_\mathrm{h}: \bnabla\testf  \dO \notag 
- \int_{\Omega_\mathrm{h}}\bnabla \cdot \bu^\mathrm{n+\alpha}_\mathrm{h} q \dO \notag \\
+&\sum_{\mathrm{e}=1}^{n_\mathrm{el}} \int_{\Omega^\mathrm{e}} \tau_m\left(\rho\bu^{\mathrm{n}+\alpha}_\mathrm{h}\cdot \bnabla \testf + \bnabla q \right) 
   \cdot \left( \rho \frac{\partial{\bu}^{\mathrm{n}+\alpha_\mathrm{m}}_\mathrm{h}}{\partial t}
+\rho \bu^{\mathrm{n}+\alpha}_\mathrm{h} \cdot \bnabla \bu^{\mathrm{n}+\alpha}_\mathrm{h}
   -\bnabla \cdot \bs^{\mathrm{n}+\alpha}_\mathrm{h}
   -\bb
   \right) \mathrm{d}\Omega^\mathrm{e} \notag \\ 
   +&\sum_{\mathrm{e}=1}^{n_\mathrm{el}} \int_{\Omega_\mathrm{e}}\bnabla \cdot \testf \tau_c \rho \bnabla \cdot \bu^\mathrm{n+\alpha}_\mathrm{h}\mathrm{d}\Omega^\mathrm{e}   
   =  \int_{\Omega_\mathrm{h}} \boldsymbol{b}
   \cdot \testf \dO
   +\int_{\Gamma_\mathrm{h}} \boldsymbol{h} \cdot \testf  \mathrm{d\Gma},
   \label{eq:NS-G-alpha}
\end{align}
\noindent where the 4th and 5th terms on the left hand side represent the stabilization terms applied on each element locally. All the other integrals constitute the Galerkin terms and they are evaluated at $t^{n+1}$.
The stabilization parameters $\tau_m$ and $\tau_c$ appearing in the element level integrals are 
the least-squares metrics, which are added to the fully discretized formulation \cite{yuri}. Refer to \cite{jaiman2016,jaiman_caf2016} for the detailed derivation of Eq.~(\ref{eq:NS-G-alpha}).


Linearization of the variational form is carried out by the standard Newton-Raphson technique. Let $\Delta \bu$ and $\Delta {p}$ denote the increment in the velocity and pressure variables. The linearized matrix form of Eq. \eqref{eq:NS-G-alpha} is written as:
\begin{align}
	\vec{M} \Delta \bu + \theta \vec{K}_d  \Delta \bu + \theta \vec{N}\Delta \bu + \theta \vec{G} \Delta {p} &= \vec{R}_m \label{eq:fluidMatrix1} \\
	- (\vec{G}_M)^T \Delta \bu- (\vec{G}_K)^T \Delta \bu + \vec{C} \Delta {p} &= R_c \label{eq:fluidMatrix2}
\end{align}
where $\vec{M}$ is the mass matrix, $\vec{K}_d$ is the diffusion matrix, $\vec{N}$ is the convection matrix, $\vec{G}$ is the pressure gradient operator. $\vec{G}_M$, $\vec{G}_K$, and $\vec{C}$ are the contribution of mass matrix, stiffness matrix and pressure matrix for the continuity equation respectively. The parameter $\theta = 2 \Delta t (1+\rho_{\infty}) / (3-\rho_{\infty})$ is a scalar, whereas $\rho_{\infty}$ denotes the spectral radius which acts as a parameter to control the high-frequency damping \cite{jaiman_caf2016}. $\vec{R}_m$ and $R_c$ are the right-hand residual vectors in the linearized form of the momentum and continuity equations, respectively.

\subsection{Matrix form of full-order flow problem}
\label{Sec:MatrixFormFOM}
We solve the N-S equations at discrete time steps to capture the transient flow characteristics, which lead to a sequence
of linear systems of equations via Newton-Raphson type iterations. 
Using Eqs. (\ref{eq:fluidMatrix1}) and (\ref{eq:fluidMatrix2}), the linearized form of the Navier-Stokes equations can be written as the coupled system of equations:
\begin{equation}
\left[ \begin{array}{cc} \bM + \theta (\bK_d+\bN) & \theta \bG \\ -(\bG_M^T+\bG_K^T) & \bC \end{array}\right ] 
\left\{\begin{array}{c}\Delta\bu \\ \Delta\bp\end{array}\right\} =
\left\{\begin{array}{c}\bR_m \\ \bR_c\end{array}\right\}
\label{eq:presvel}
\end{equation}
This coupled matrix form is complicated to solve since there are two sets of pressure and velocity modes, whereby the velocity modes must be captured accurately before evaluating the pressure modes.
Linear solvers are needed for the solution of momentum  and the pressure equations.
We first solve the symmetric matrix of the pressure projection using the Conjugate Gradient (CG), with a diagonal preconditioner \cite{barrett1994templates}.  This provides a set of projection
vectors for the low-pressure modes of 
the coupled Navier-Stokes system.
Note that to solve this equation we do not need the left-hand-side matrix in the explicit form, but instead we perform the required matrix-vector products.
Without disturbing the projected low-pressure modes, we then use the standard Generalized Minimal Residual (GMRES) solver based on the modified Gram-Schmidt orthogonalization \cite{saad1986} for the velocity and pressure increments using Eq.~(\ref{eq:presvel}) within the Newton-Raphson loop. The GMRES is an iterative Krylov-subspace method, which minimizes the residual norm of all vectors in the Krylov subspace. The Arnoldi recurrence generates a sequence of orthogonal vectors that span the Krylov space \cite{saad1986} and 
the solution to the least squares system provides an approximate solution by orthogonal projection.
A Krylov cycle involves the equal number of steps of the Arnoldi recurrence and matrix-vector products for the Krylov space. 
Each new Arnoldi vector is explicitly orthogonalized against all previous vector, which increases the storage linearly with the iteration count until convergence and the work increases quadratically with the iteration count.
%
In the above full-order iterative solution, the linear equation solution via the Krylov-subspace has a robust performance and convergence rate for the steepest descent. However, such matrix-free Krylov methods for the full-order model require much more storage of intermediate results, and they are generally non-local. Furthermore, the formations of finite element terms are also computationally expensive processes for element-by-element integration in the full-order analysis using Eq.~(\ref{eq:NS-G-alpha}) at each iteration, 
which can be replaced by some equation-free 
approximation via a deep neural network.
We next present our data-driven method based on deep-learning and the stochastic gradient descent approach.
%
%
%
\section{Model Reduction via Convolutional Neural Network}
\label{CNN}
Deep learning utilizes multiple layers of representation to learn relevant features automatically from training data via deep neural networks. These deep neural networks can be considered as 
multi-iterative coarse-graining scheme which provides a successive 
process to learn  relevant dynamics from the data by combining low-level 
features with higher-level features.
In this study, we consider the convolutional neural network for our deep learning technique to extract relevant features from fluid dynamics data and then predict the fluid forces.
With this objective, we hypothesize that a properly trained CNN-based deep-learning technique can find the mapping between flow parameters (e.g.,  force coefficients, Strouhal number) and the design parameters (e.g., aspect ratio, rounding angle, angle of attack) of interest and hence can be used for predictions.
The training phase of the CNN comprises the input function, 
the feed-forward process, and the back-propagation process. We feed the input function as a matrix of values derived from a unique mapping of the design parameter.
Then 
the feed-forward process initially assumes the features of the wake flow and 
performs four operations on this input matrix:
(i) convolution: a linear process which extracts the local features of the 
flow using small matrices called kernels or filters 
and sliding-window operations, 
(ii) nonlinear rectification: adds non-linearity to the outcome of the convolution, 
(iii) down-sampling (if required): reduces the memory consumption 
of the process at the expense of data dropout,
and (iv) fully connected layers: connect all the extracted features 
to output the wake flow parameter. 

In a deep neural network, input features are mapped or transformed into hidden units to form 
new features to feed to the next layer. 
While the terminology of feature map is quite broad in the field of machine learning, 
a feature map is essentially a function which maps the input data vector to feature space 
and provides the encoding of features
and their locations for convolutional layers.
Using the discrepancy between the FOM and the predicted result, the back-propagation 
process corrects the assumed convolution kernel values and weights 
of the fully connected layer. 
By successively performing feature extraction, the networks learn to emphasize dynamically relevant features while having a diminishing effect on the irrelevant features.
After a prescribed number of these iterations,
the CNN-based learning is deemed to be properly trained and ready to predict the force 
coefficients of the new geometry set. 
The components of the iterative training are elaborated as follows 
for our CNN-based deep learning.

\subsection{Input function}
The CNN starts with an input function which maps the macroscopic parameter of concern with the independent variable(s). For the sake of explanation let the input function be a matrix $D(x) = D_{ij}$ of the size $\mathbb{R}^{m \times n}$. It is critical to define the input function for a large enough domain with a proper refinement. However, we show in Section \ref{Results} that the refinements required for CNN are much less than the finite element NS solvers.
After generating the input function, the feed-forward process is applied to it to extract the dominant features of the fluid flow.

\subsection{Feed-forward process}
\label{sec:Feedforward}
The feed-forward process applies the operations: convolution, rectification, and down-sampling in sequence, 
first on the input matrix, then on the output obtained from the above operations. 
We refer to each of these operations as a layer and the combination of 
adjacent convolution, rectification and down-sampling as a composite layer. 
We denote the output of each layer of the $l^{th}$ 
composite layer as $Y^{Cl}, Y^{Rl}$ and $Y^{Dl}$, respectively. 
The total number of composite layers, denoted by $N$, is a critical factor 
influencing the accuracy of the entire convolutional network. 
Such system variables are referred to as hyper-parameters and their values 
need to be determined for an optimal performance.
The feed-forward process takes a set of 2D vectors as input and applies 
the discrete convolution operation with a number of 2D kernels. 
For the first convolution operation, the input is the input 
function: $D(x)$. Let us denote the total number of kernels in the $l^{th}$ 
composite layer by $k_l$ and group all those kernels with size $\delta \times \xi$ into the tensor 
$K^l \in \mathbb{R}^{\delta \times \xi \times k_l}$. When properly trained, 
each of these kernels represents a feature of the system contributing 
to the variation of the unknown output parameter due to the change in the design parameter. 
The convolution kernel size
determines the input neighborhood size which 
represents a particular feature. 

When we apply the first layer of 2D convolution on the input matrix, 
i.e. $(l=1)$, it outputs the 3D tensor $Y^{C1}=\{Y^{C1}_{ijk}\}$ \citep{klette2014concise}:
\begin{equation}
\label{Eq:Convolution}
Y_{ijk}^{C1} = D_{ij} \star K^1_{ijk} = \sum_{b=1}^{n}\sum_{a=1}^{m} D_{ab} K^1_{(i-a+1)(j-b+1)k}, \: k=1,2,...,k_1,  
\end{equation}
where the symbol $\star$ denotes the convolution operation, which allows extracting local features with some prior knowledge.
Eq. (\ref{Eq:Convolution}) changes slightly if the convolution blocks 
are skipped on more than one element of the input matrix on any direction. 
The skipping lengths along the two directions of the input, termed as the stride $s_l = [s_{l1}\:\: s_{l2}]$ is also an important hyper-parameter.
Except for the first composite layer, the input to the convolution layer is 
a 3D tensor. The $l^{th}$ layer takes each 2D slice of the tensor $Y^{D(l-1)}$ 
and convolutes them with all the kernels which create a 3D tensor $Y^{Cl}$ 
with a depth of $k_{(l-1)}\times k_l$. These additional convolutions 
increase the neighborhood size of a particular feature and may cause some errors as well. 

Since convolution is a linear operation, it is required to introduce a nonlinear 
rectification process to capture the complex nonlinear flow features 
such as the bluff body wake. We employ the common rectifier linear unit (reLU) for 
this purpose which yields $Y^{Rl}=\{Y^{Rl}_{ijk}\}$ as:
\begin{equation}
Y_{ijk}^{Rl} = \max (Y_{ijk}^{Cl},0).
\end{equation}
{The reLU is generally observed to be a faster learning non-linearization process than the other commonly used $\tanh(x)$ and sigmoid $(1+e^{-x})^{-1}$ functions \cite{krizhevsky2012imagenet}.}
By generalizing the convolution for the $l^{th}$ layer, 
the size $v_l$ of $Y^{Cl}$ will be: $v_l=\ceil*{\frac{m}{\prod s_{l1}}} 
\times \ceil*{\frac{n}{\prod s_{l2}}} \times \prod k_l$ which gets larger 
with smaller strides and a large number of kernels and convolution layers.
A down-sampling/pooling layer between convolutions can be used to 
reduce the size of the CNN matrices at the cost of losing some data. {The commonly used pooling techniques are the maximum, average and minimum where the respective function (max, avg or min) is applied to small windows of the rectified output, generally with strides equal to the pooling window size. It reduces the size of the feature map and also drops some data.}
%
After the $N$ composite layers, the feed-forward process ends with an additional $1D$ 
vector called a fully connected layer which has the size 
of $\mathbb{R}^{\beta \times 1}$ where $\beta$ is the stacked size of $v_N$.
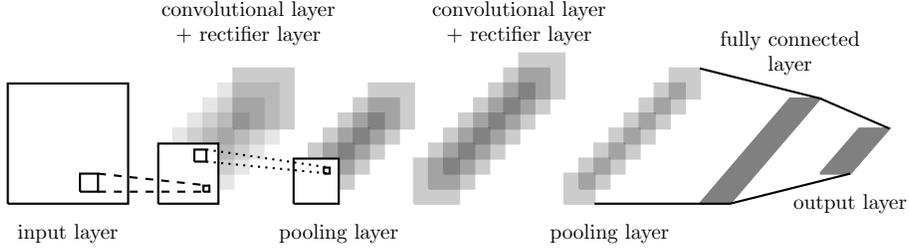
\begin{figure}[t!]
	\centering
	\begin{tikzpicture}[thick,scale=0.8, every node/.style={scale=0.8}]
		\node at (0,-0.5){\begin{tabular}{c}input layer \end{tabular}};
		
		\draw (-1,0) -- (1,0) -- (1,2) -- (-1,2) -- (-1,0);
		\draw (0.2,0.2) -- (0.5,0.2) -- (0.5,0.5) -- (0.2,0.5) -- (0.2,0.2);
		
		\node at (3,3){\begin{tabular}{c}convolutional layer\\+ rectifier layer\\\end{tabular}};
		
		\draw[fill=black,opacity=0.2,draw=black] (2.75,1.25) -- (3.75,1.25) -- (3.75,2.25) -- (2.75,2.25) -- (2.75,1.25);
		\draw[fill=black,opacity=0.1,draw=black] (2.5,1) -- (3.5,1) -- (3.5,2) -- (2.5,2) -- (2.5,1);
		\draw[fill=black,opacity=0.1,draw=black] (2.25,0.75) -- (3.25,0.75) -- (3.25,1.75) -- (2.25,1.75) -- (2.25,0.75);
		\draw[fill=black,opacity=0.1,draw=black] (2,0.5) -- (3,0.5) -- (3,1.5) -- (2,1.5) -- (2,0.5);
		\draw[fill=black,opacity=0.1,draw=black] (1.75,0.25) -- (2.75,0.25) -- (2.75,1.25) -- (1.75,1.25) -- (1.75,0.25);
		\draw[fill=white,opacity=1,draw=black] (1.5,0) -- (2.5,0) -- (2.5,1) -- (1.5,1) -- (1.5,0);
		\draw (2.25,0.2) -- (2.35,0.2) -- (2.35,0.3) -- (2.25,0.3) -- (2.25,0.2);
		\draw[dashed] (0.5,0.5) -- (2.35,0.3);
		\draw[dashed] (0.5,0.2) -- (2.35,0.2);
		\draw (2.1,0.7) -- (2.3,0.7) -- (2.3,0.9) -- (2.1,0.9) -- (2.1,0.7);
		
		\node at (4.5,-0.5){\begin{tabular}{c}pooling layer\\\end{tabular}};
		
		\draw[fill=black,opacity=0.2,draw=black] (5,1.25) -- (5.75,1.25) -- (5.75,2) -- (5,2) -- (5,1.25);
		\draw[fill=black,opacity=0.2,draw=black] (4.75,1) -- (5.5,1) -- (5.5,1.75) -- (4.75,1.75) -- (4.75,1);
		\draw[fill=black,opacity=0.2,draw=black] (4.5,0.75) -- (5.25,0.75) -- (5.25,1.5) -- (4.5,1.5) -- (4.5,0.75);
		\draw[fill=black,opacity=0.2,draw=black] (4.25,0.5) -- (5,0.5) -- (5,1.25) -- (4.25,1.25) -- (4.25,0.5);
		\draw[fill=black,opacity=0.2,draw=black] (4,0.25) -- (4.75,0.25) -- (4.75,1) -- (4,1) -- (4,0.25);
		\draw[fill=white,opacity=1,draw=black] (3.75,0) -- (4.5,0) -- (4.5,0.75) -- (3.75,0.75) -- (3.75,0);
		\draw (4.25,0.5) -- (4.35,0.5) -- (4.35,0.6) -- (4.25,0.6) -- (4.25,0.5);
		\draw[dotted] (2.3,0.9) -- (4.35,0.6);
		\draw[dotted] (2.3,0.7) -- (4.35,0.5);
		
		\node at (7.5,3){\begin{tabular}{c}convolutional layer\\+ rectifier layer\\\end{tabular}};
		
		\draw[fill=black,opacity=0.2,draw=black] (7.5,1.75) -- (8.25,1.75) -- (8.25,2.5) -- (7.5,2.5) -- (7.5,1.75);
		\draw[fill=black,opacity=0.2,draw=black] (7.25,1.5) -- (8,1.5) -- (8,2.25) -- (7.25,2.25) -- (7.25,1.5);
		\draw[fill=black,opacity=0.2,draw=black] (7,1.25) -- (7.75,1.25) -- (7.75,2) -- (7,2) -- (7,1.25);
		\draw[fill=black,opacity=0.2,draw=black] (6.75,1) -- (7.5,1) -- (7.5,1.75) -- (6.75,1.75) -- (6.75,1);
		\draw[fill=black,opacity=0.2,draw=black] (6.5,0.75) -- (7.25,0.75) -- (7.25,1.5) -- (6.5,1.5) -- (6.5,0.75);
		\draw[fill=black,opacity=0.2,draw=black] (6.25,0.5) -- (7,0.5) -- (7,1.25) -- (6.25,1.25) -- (6.25,0.5);
		\draw[fill=black,opacity=0.2,draw=black] (6,0.25) -- (6.75,0.25) -- (6.75,1) -- (6,1) -- (6,0.25);
		\draw[fill=black,opacity=0.2,draw=black] (5.75,0) -- (6.5,0) -- (6.5,0.75) -- (5.75,0.75) -- (5.75,0);
		
		\node at (9,-0.5){\begin{tabular}{c}pooling layer\\\end{tabular}};
		
		\draw[fill=black,opacity=0.2,draw=black] (10,1.75) -- (10.5,1.75) -- (10.5,2.25) -- (10,2.25) -- (10,1.75);
		\draw[fill=black,opacity=0.2,draw=black] (9.75,1.5) -- (10.25,1.5) -- (10.25,2) -- (9.75,2) -- (9.75,1.5);
		\draw[fill=black,opacity=0.2,draw=black] (9.5,1.25) -- (10,1.25) -- (10,1.75) -- (9.5,1.75) -- (9.5,1.25);
		\draw[fill=black,opacity=0.2,draw=black] (9.25,1) -- (9.75,1) -- (9.75,1.5) -- (9.25,1.5) -- (9.25,1);
		\draw[fill=black,opacity=0.2,draw=black] (9,0.75) -- (9.5,0.75) -- (9.5,1.25) -- (9,1.25) -- (9,0.75);
		\draw[fill=black,opacity=0.2,draw=black] (8.75,0.5) -- (9.25,0.5) -- (9.25,1) -- (8.75,1) -- (8.75,0.5);
		\draw[fill=black,opacity=0.2,draw=black] (8.5,0.25) -- (9,0.25) -- (9,0.75) -- (8.5,0.75) -- (8.5,0.25);
		\draw[fill=black,opacity=0.2,draw=black] (8.25,0) -- (8.75,0) -- (8.75,0.5) -- (8.25,0.5) -- (8.25,0);
		
		\node at (12,2.5){\begin{tabular}{c}fully connected \\layer\\\end{tabular}};
		
		\draw[fill=black,draw=black,opacity=0.5] (10.5,0) -- (11,0) -- (12.5,1.75) -- (12,1.75) -- (10.5,0);
		\draw (10.5,2.25) -- (12.5,1.75);
		\draw (8.75,0) -- (11,0);
		
		\node at (13,0){\begin{tabular}{c}output layer \end{tabular}};
		\draw (13.65,1.25) -- (12.5,1.75);
		\draw (13,0.5) -- (11,0);
		
		\draw[fill=black,draw=black,opacity=0.5] (12.5,0.5) -- (13,0.5) -- (13.65,1.25) -- (13.15,1.25) -- (12.5,0.5);
	\end{tikzpicture}
	\caption{An abstract representation of a convolutional neural network (CNN) architecture, which illustrates the alternative sequential stages between convolutional and pooling layers. While the convolution step is denoted in dashed lines, the pooling-layer window is denoted in dotted lines. Units in a convolutional
layer are feature maps and each unit
is connected to local patches in the feature maps of the previous layer through a set of weights.
    }
	\label{fig:CNN_Schematic}
\end{figure}
Figure \ref{fig:CNN_Schematic} summarizes a general architecture of the convolutional neural network.
In the figure, a convolutional layer represents two sub-layers namely convolution and rectifier. The feature maps of the final downsampling layer are fed into the fully connected layers for the feed-forward process.
While the convolutional layer plays a role to detect local connections of features from the previous layer, the pooling
layer is to collapse the similar features into one.
The final set of predictions for the force coefficient of the feed-forward 
process $C_{CNN}^F$ for a training set with $S$ inputs is given by:
\begin{equation}
[C_{CNN}^F]^{S\times 1} = [w]^{S \times \beta} [Y^N]^{\beta \times 1}.
\end{equation}
where $w$ is the weight matrix of the fully connected layer. The values of the kernels and these weights are the adjustable parameters of the system and denote them as a set $W$ for the ease of explanation. $C_{CNN}^F$ is compared with the full order result ($C_{FOM}^F$) and the weights are then adjusted to minimize the error using the back-propagation process.
As being a universal black-box approximation, the above neural networks with convolution layers can be effective to deal with the local interaction and multiscale nature of Navier-Stokes PDE system.
In subsection \ref{sec:physics_analogy}, we provide a connection between the convolution process with the memory kernels in the Mori-Zwanzig formalism 
used for the model reduction of a full-order hydrodynamic system.
\subsection{Back-propagation process}
The role of back-propagation is to train a multi-layered neural network to learn the relevant features from any arbitrary input-output datasets.
The training phase is an iterative process, which continuously minimizes the 
error between the predicted and target outputs. 
It compares the output of the feed-forward pass with the full-order result 
and corrects the set of weights ($W$) to minimize the error.
While the network is initialized with randomly chosen weights, the backpropagation process attempts to find a local minimum of the error function.
Let us represent the entire feed-forward process for the $p^{th}$ 
input matrix ($d^p$) with the function $H(d^p,W)$. We define a cost 
function $G$ to measure the discrepancy between the feed-forward 
prediction and full-order fluid coefficient:
\begin{equation}
E^{p} = G(C^F_{FOM},H(d^p,W)).
\end{equation}
In this study, the cost function $G$ is the root mean square error $L_2$ function. 
Now the target is to update the weight set $W$ to minimize the error $E^p$ using a standard gradient descent back-propagation method.

For clarity, we denote the layer number in superscripts and the iteration number in subscripts unless otherwise mentioned. 
For simplicity let us denote the output of the $l^{th}$ composite layer of the 
feed-forward process as $Y^l$. It can be related to the previous layer 
output $Y^{(l-1)}$ and the weights of the $l^{th}$ layer $W^l$:
\begin{equation}
Y^l = F(W^l,Y^{(l-1)}),
\end{equation}
where $F$ represents a single pass of convolution, rectification, and down-sampling. 
Note that $Y^0=D$ and $Y^{(N+1)}=C_{CNN}^F$. 
The back-propagation process starts at this predicted value where the error gradient 
of the fully connected layer can be determined first. 
We next calculate the error gradient with respect to the weights and the
previous layer output recurrently in the backward direction. When the error gradient of the $l^{th}$ output layer: $\frac{\partial E^p}{\partial Y^l}$ is known, 
we can get the error gradients using the chain rule:
\begin{equation}
\begin{array}{cc}
\displaystyle \frac{\partial E^p}{\partial W^l} = \frac{\partial F}{\partial W}\left(W^l,Y^{(l-1)}\right)\frac{\partial E^p}{\partial Y^l},\\[8pt]
\displaystyle
\frac{\partial E^p}{\partial Y^{(l-1)}} = \frac{\partial F}{\partial Y}\left(W^l,Y^{(l-1)}\right)\frac{\partial E^p}{\partial Y^l},
\end{array}
\end{equation}
where $\frac{\partial F}{\partial W}$ and $\frac{\partial F}{\partial Y}$ are the Jacobians of $F$ relative to $W$ and $Y$, and are evaluated at the $l^{th}$ layer. 
The successive gradients during the backpropagation can also be linked with the variational least-square minimization.
After the evaluation of all the error gradients  $\frac{\partial E^p}{\partial W}$, we use the stochastic gradient descent method with momentum (SGDM) \citep{rumelhart1988learning} to adjust the parameter set for the $T^{th}$ iteration:
\begin{equation}
W_T = W_{T-1}-\gamma \frac{1}{S} \sum_{p=1}^S \frac{\partial E^p}{\partial W}+\eta (W_{T-1}-W_{T-2})
\label{Eq:Sgdm}
\end{equation}
where $\gamma>0$ is the learning rate which is also a hyper-parameter discussed in Section \ref{HyperP}. $\eta \in [0,1]$ is called the momentum and is the hyper-parameter which 
determines the contribution from the previous gradient correction. 
The gradient in the SGDM is an expectation, which 
may be approximately estimated using a small set of samples. 
Refer to \citep{lecun2012efficient} for the proof of convergence and the derivation of the SGDM method. In the next section, we provide the linking between the deep kernel learning with the momentum term and the simplified analytical form 
of the Navier Stokes momentum equation.
We have outlined the detailed formulation of our CNN-based deep learning algorithm via stochastic gradient descent.
The present approach is solely data-driven and does not 
rely explicitly on the underlying governing equations of the dynamical system.

\subsection{Physical analogy}
\label{sec:physics_analogy}
Both flow physics and deep learning rely on many degrees of freedoms, which interact in a nonlinear and multiscale manner.
Here, we attempt to demonstrate a physical analogy for deep learning in the context of Navier-Stokes equations for the fluid flow modeling.
%
This analogy will provide a fundamental basis and a conceptual underpinning to understand how the deep learning
can provide a useful reduced model for flow dynamics.

The deep learning based on convolutional neural networks can be considered as a black box mapping between two functions \cite{raissi2017machine}:
\begin{equation}
y(\xx) = \mathcal{L}^{\phi}F(\xx),
\label{eq: blackbox}
\end{equation}
where $y$ and $F$ are the mappings of the input and output with the independent variable $\xx$. The aim of CNN process is to determine the unknown mapping $\mathcal{L}$ by extracting its features $\phi$ using available data. Moreover, the CNN transforms this problem to determine some unknown kernels $\kappa$ in the equation:
\begin{equation}
\mathcal{L}^{\phi}F(\xx) = (\kappa\star F)=\int \kappa(\xx-\xx')F(\xx')d\xx',
\label{Eq: kernelCNN}
\end{equation}
where $\xx'$ is a dummy variable.
The task of CNN is now to estimate these kernels (memory effect) by a learning process based on available input-output combinations obtained by full-order methods. 
Essentially, the CNN-based process facilitates the systematic 
approximations of the memory effects during the model reduction. 
We will justify this approximation property of deep learning by linking with the Mori-Zwanzig formalism.
Before we provide an analogy of the convolution effect and the stochastic iterative correction process, we transform the incompressible Navier Stokes momentum equation into a simplified integral-differential form.
\subsubsection{Transformation of the Navier-Stokes equation} 
To begin,
let $\phi$ be the divergence-free velocity potential function in Eq. (\ref{eq:NS}), such that $\boldsymbol{u} = \boldsymbol{\nabla}\phi$. Substituting this for a very small volumetric fluid region ($\Delta V$), this gives:
\begin{align}
\frac{\partial \boldsymbol{\nabla}\phi}{\partial t} + (\boldsymbol{\nabla}\phi \cdot \boldsymbol{\nabla}) \boldsymbol{\nabla}\phi = -\frac{1}{\rho}\boldsymbol{\nabla}p + \nu \boldsymbol{\nabla^2}\boldsymbol{\nabla}\phi. \label{eq:VelPot}
\end{align}
which can be reduced to the reaction-diffusion equation:
\begin{align}
\frac{\partial\psi}{\partial t} - \nu{\nabla^2}\psi = \frac{\Delta p}{2\mu}\psi
\label{eq:Reac-diffCNN}
\end{align}
where $\psi=-2\nu\ln{\phi}$. 
Detailed steps of the transformation 
of the Navier-Stokes momentum equation 
into the reaction-diffusion equation 
is presented in Appendix A. 
From a statistical viewpoint, Eq.~(\ref{eq:Reac-diffCNN}) is a particular case of the Einstein-Kolmogorov differential equation:
\begin{align}
\frac{\partial \Upsilon}{\partial t} = -\frac{\partial A\Upsilon}{\partial x} + \frac{\partial^2 B\Upsilon}{\partial x^2},
\label{eq:EKde}
\end{align}
where $\Upsilon(\xx,t;\xx_0,t_0)$ is the concentration at spatial point $\xx$ at time $t$ of particles which started undergoing Brownian motion at time $t_0$ from point $\xx_0$. Hence $\Upsilon(\xx,t;\xx_0,t_0) \cdot \Delta V$ is the probability of finding a particle undergoing Brownian motion. 
The connection between the partial differential equations and the Brownian Motion
can be constructed by the basic differentiation
of distribution function of the underlying Gaussian stochastic process.
A statistical-mechanical theory of many fluid particle system can be employed to link the 
collective transport and the Brownian motion. 
The following equation so-called the Einstein-Kolmogorov equation relates the particle concentration with the initial concentration:
\begin{align}
\Upsilon(\xx,t;\xx_0,t_0) = \int \Upsilon(\xx,t;P,\theta)\Upsilon(P,\theta;\xx_0,t_0)\,dV_P,
\end{align}
where $t_0 <\theta < t$ and the integral is taken all over the spatial domain. This equation leads to Eq. (\ref{eq:EKde}) as shown in \cite{tikhonov2013equations}. Eq. (\ref{eq:EKde}) relates the stochastic process of the Brownian motion to a continuum partial differential equation on concentration. 

Next, we can formulate a general solution for the reaction-diffusion equation, Eq. (\ref{eq:Reac-diffCNN}) using the Green's function.
Let $\mathcal{L}$ be the operator $\mathcal{L} = \frac{\partial}{\partial t}-\nu\boldsymbol{\nabla^2}$. The differential equations for $\psi(\xx,t)$ and the Green's function, $G(\xx,t;\boldsymbol{\xi},\tau)$ for $\xx,\boldsymbol{\xi}\in \mathcal{D}$ and $t,\tau \geq 0$ can be taken as:
\begin{align}
\mathcal{L}\psi(\xx,t) = \frac{\Delta p}{2\mu}\psi(\xx,t),\\
\mathcal{L}G(\xx,t;\boldsymbol{\xi},\tau)=\delta (\xx-\boldsymbol{\xi})\delta(t-\tau),
\end{align}
where $\mathcal{D}$ is the fluid domain of interest and $\boldsymbol{\xi}$ and $\tau$ are dummy variables.
This leads to the integral form of the Green's function based general solution for non-homogeneous diffusion equation \cite{haberman1983elementary} (see Appendix B):
\begin{equation}
\begin{split}
\psi(\xx,t)= \int_0^{\infty}\int_V G\frac{\Delta p}{2\mu}\psi(\xx,t)\, dV_{\boldsymbol{\xi}}d\tau +\int_V G\psi(\xx,0)\,dV_{\boldsymbol{\xi}} \\ +\nu \int_0^{\infty} \oint_S \left[G\boldsymbol{\nabla}_{\boldsymbol{\xi}} \psi  +\psi \boldsymbol{\nabla}_{\boldsymbol{\xi}} G \right]\boldsymbol{n}\,dSd\tau.
\label{eq:nonhomogCNN}
\end{split}
\end{equation}
Note that the second and third terms represent the initial and boundary conditions. Considering only the first term, Green's function is the unknown kernel analogous to the kernels of the Eq. (\ref{Eq: kernelCNN}).
Then the blackbox process $\mathcal{L}^{\phi}$ in Eq. (\ref{eq: blackbox}) is analogous to the fixed point iteration of the nonlinear functional map of Eq. (\ref{eq:nonhomogCNN}) given by $\mathcal{F}$:
\begin{equation}
\psi_{m+1}(\xx,t) = \mathcal{F}\left(\psi_m(\xx,t)\right),
\end{equation}
where $m$ is the iteration number. 
The above steps provide a physical process to construct the direct mapping function from the Navier-Stokes momentum equation for the hydrodynamic variable of interest. 
This integral transformation of the variable of interest can also be realized using the deep architecture where the weights between the successive layers are connected through a mapping function. 
%
%
We next present the Mori-Zwanzig formalism to establish a connection 
of the reduced system obtained from the Navier-Stokes equation.
During the coarse-grained representation via the Mori-Zwanzig formalism, the complexity of large degrees of freedom in the full-order model is replaced by the convolution (memory) kernels.
In the deep learning, the CNN process provides systematic approximations for the memory effects of the original dynamics.

\subsubsection{Mori-Zwanzig formalism and convolution integral}
We briefly discuss the Mori-Zwanzig formalism \cite{mori1965transport, zwanzig1973nonlinear} in the context of modeling memory effects during the CNN-based deep learning. 
This formalism provides a strategy for integrating out a subset of variables in a generic dynamical system through the 
coarse-graining process.
Let the coarse-grained modes of the function $\psi$ in Eq. (\ref{eq:Reac-diffCNN}) be $\chi$ such that $\psi(\chi_0,t)=g(\chi)$, where the initial condition $\chi(0) = \chi_0$ with $\chi_0 \in L^2$. Consider the decomposition of $\psi$ as the slow (resolved) modes $\widehat{\psi}=\psi(\widehat{\chi}_0,t)$ and the fast (unresolved) modes $\widetilde{\psi}=\psi(\widetilde{\chi}_0,t)$. 
%
%
As presented in Appendix C, the Mori-Zwanzig equation for the resolved variable of interest $\widehat{\psi}$ can be derived as:
\begin{equation}
\frac{\partial \widehat{\psi}}{\partial t}=\underbrace{\Theta\widehat{\psi}}_\text{Markovian} + \underbrace{\int_0^t K(t_1)\widehat{\psi}(t-t_1)\,dt_1}_\text{Convolution} + \underbrace{F(\widehat{\chi}_0,t)}_\text{Fluctuation}
\label{eq:MZFormalismCNN}
\end{equation}
The first term in the right-hand side is a Markovian term given by  $\Theta\widehat{\psi} = e^{Lt}PL\widehat{\psi}_0$, where $P$ is a projection of $\psi$ to the slow variables $\widehat{\psi}$ and the differential operator $L$ is given by $L=(\frac{\Delta p}{2\mu}+\nu\nabla^2)$. The Markovian term depends only on the value of $\widehat{\psi}$ at time $t$. The second term depends on every value of $\psi$ during interval $[0,t]$, hence represents the memory effect via convolution process. The convolution (memory) kernel is given by $K(t){\psi} = PLe^{QLt}QL\psi$, where $Q=I-P$ is an orthogonal subspace of $P$ and $I$ denotes the identity operator. The projection operators $P$ and $Q$ are self-adjoint and they are orthogonal to each other. 
The third term  $F(\widehat{\chi}_0, t) = e^{QLt}QL\widehat{\psi}_0$ represents the fluctuation (noise) and it satisfies the orthogonal dynamics equation.
In a standard stochastic differential equation, the time evolution of the resolved variable is considered as the combination of the real-time Markovian term and the fluctuation. However, the Mori-Zwanzig formalism contains the convolution kernel with memory effect, which is not correlated with the fluctuation. The CNN-based learning process is specifically trained  to extract the memory-dependent dynamics via data-driven modeling.
In a broader sense, for fluid dynamics, we can apply this dimensionality reduction approach to build subgrid models for turbulence via data-driven approximations.
Next, we show a link between the stochastic gradient descent with the momentum and the discrete counterpart of the reaction-diffusion system.

\subsubsection{Reaction-diffusion system and stochastic gradient descent}
By considering the time-discretized form of the reaction-diffusion equation Eq. (\ref{eq:Reac-diffCNN}) for two adjacent time steps, we get:
\begin{align}
\frac{\psi_T-\psi_{T-1}}{\Delta t} = \frac{\Delta p}{2\mu}\psi_T + \nu {\nabla^2}\psi_T \\
\frac{\psi_{T-1}-\psi_{T-2}}{\Delta t} = \frac{\Delta p}{2\mu}\psi_{T-1} + \nu {\nabla^2}\psi_{T-1}
\end{align}
Subtracting the equations and rearranging the terms, we obtain
\begin{align}
\psi_T = \psi_{T-1}+\frac{2\Delta t \mu \nu}{2\mu-\Delta p \Delta t}\left({\nabla^2}\psi_T - {\nabla^2}\psi_{T-1}\right) + \frac{2\mu}{2\mu-\Delta p \Delta t}\left(\psi_{T-1}-\psi_{T-2}\right).
\end{align}
This is analogous to the stochastic gradient descent method with momentum given in Eq. (\ref{Eq:Sgdm}). The second term of the right-hand side is equivalent to the error gradient and the third term represents the momentum term.  
{The above semi-discrete form of the simplified Navier-Stokes momentum equation provides a connection with the discrete kernel learning (i.e., black-box integrator) from the stochastic gradient with momentum.}
%
%
We next present the sensitivity and convergence study 
of the proposed CNN-based stochastic gradient 
method with the momentum term.

\section{Results}
In this work, for the first time, we apply a deep convolutional network for 
the fundamental fluid dynamic problem of the flow past bluff bodies. In particular, 
we quantify the variation of force coefficients e.g., lift ($C_L$) and drag ($C_D$) as 
a function of the bluff body geometry. 
Without loss of generality, 
we consider two of the most widely studied bluff body geometry 
variations: the aspect ratio ($AR$) of an elliptical body and the rounding angle  
($\varphi$) of a square shaped body. 
Using a trained deep neural net, we will examine how the deep nets can predict well on unseen configurations (i.e., generalization property of deep nets).
\subsection{Problem setup}
\begin{figure}[h]
\centering
\begin{subfigure}{0.19\textwidth}
\centering
\includegraphics[scale=0.275]{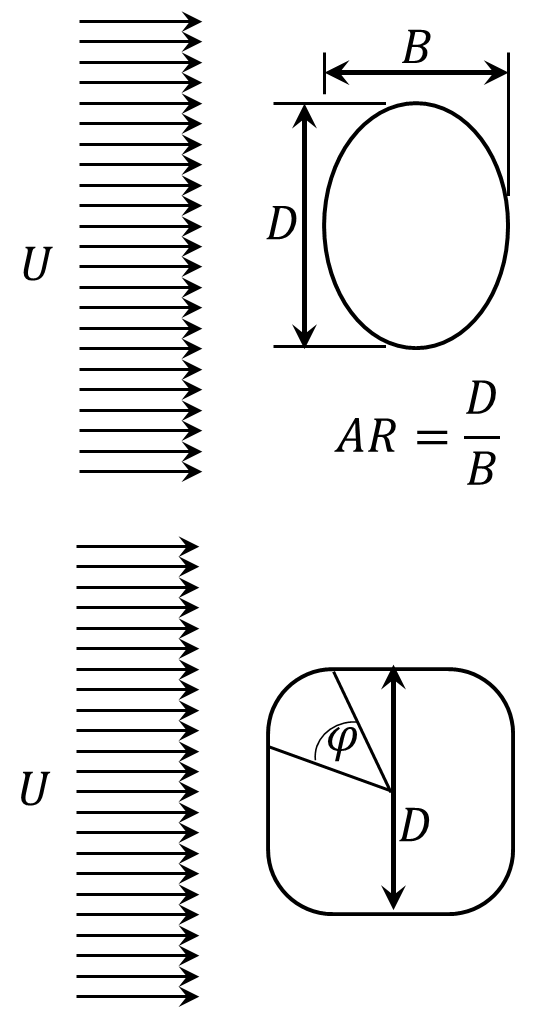}
\caption{}
\end{subfigure}~\hspace{3pt}
\begin{subfigure}{0.35\textwidth}
\centering
\dbox
{\includegraphics[scale=0.36]{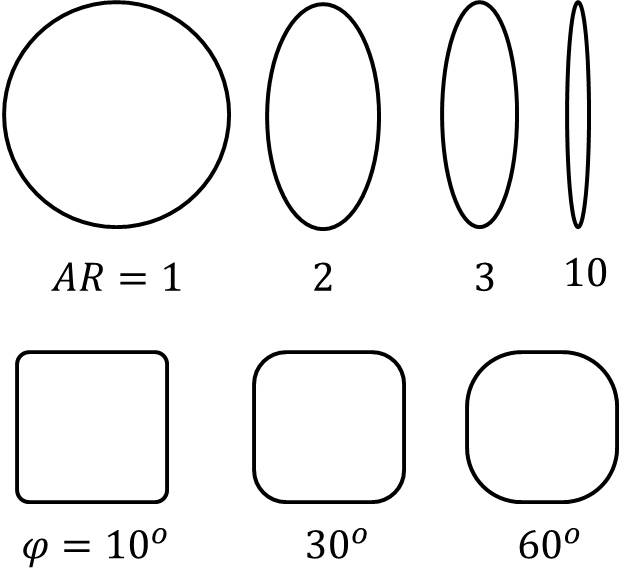}}
\caption{}
\end{subfigure}~
\begin{subfigure}{0.475\textwidth}
\centering
\includegraphics[trim={0.75cm 10.75cm 0 0},clip,scale=0.245]{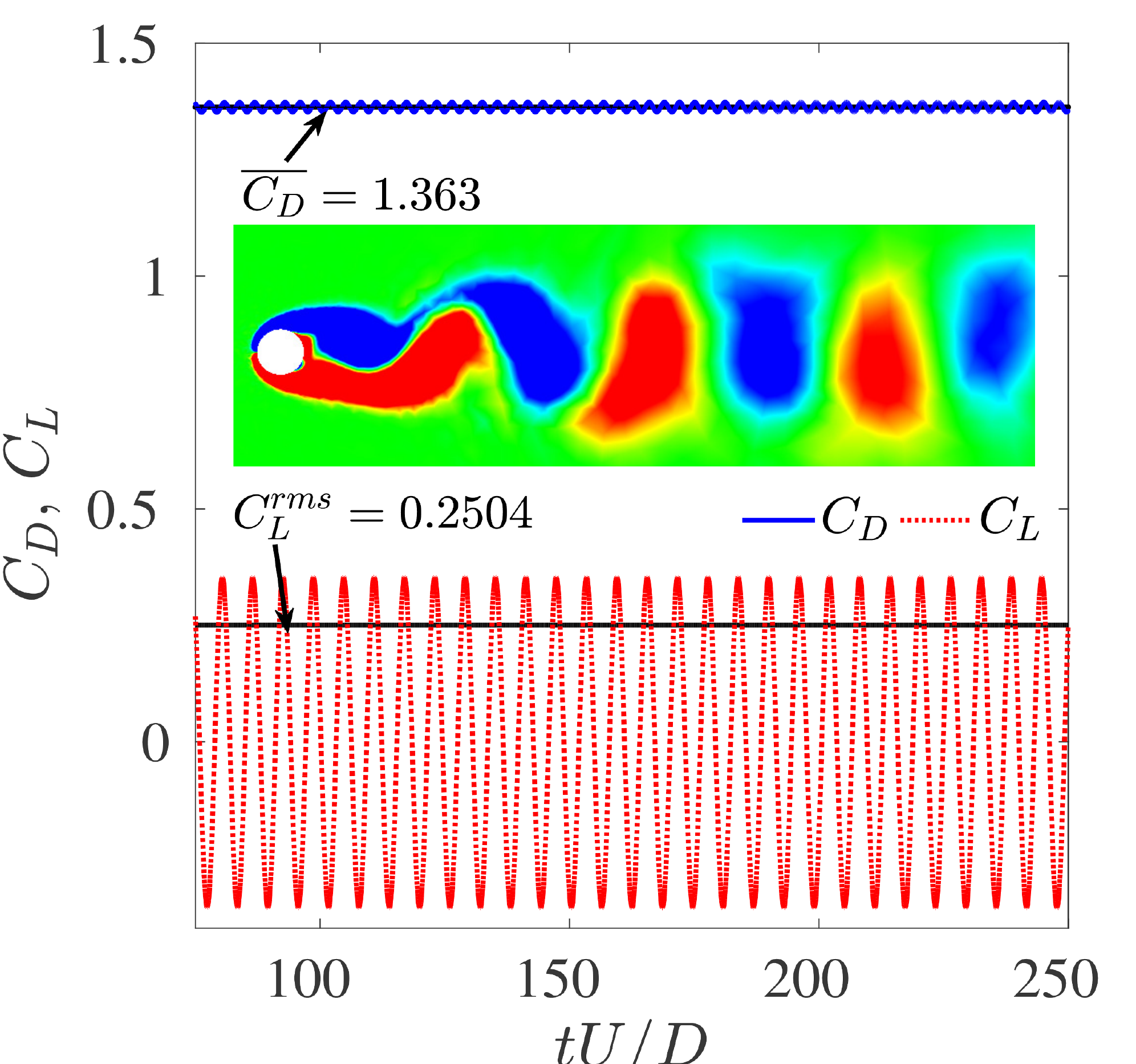}
\centering
\caption{}
\end{subfigure}\\
\begin{subfigure}{\textwidth}
\centering
\dbox
{\includegraphics[trim={0 0 0 0},clip,scale=0.39]{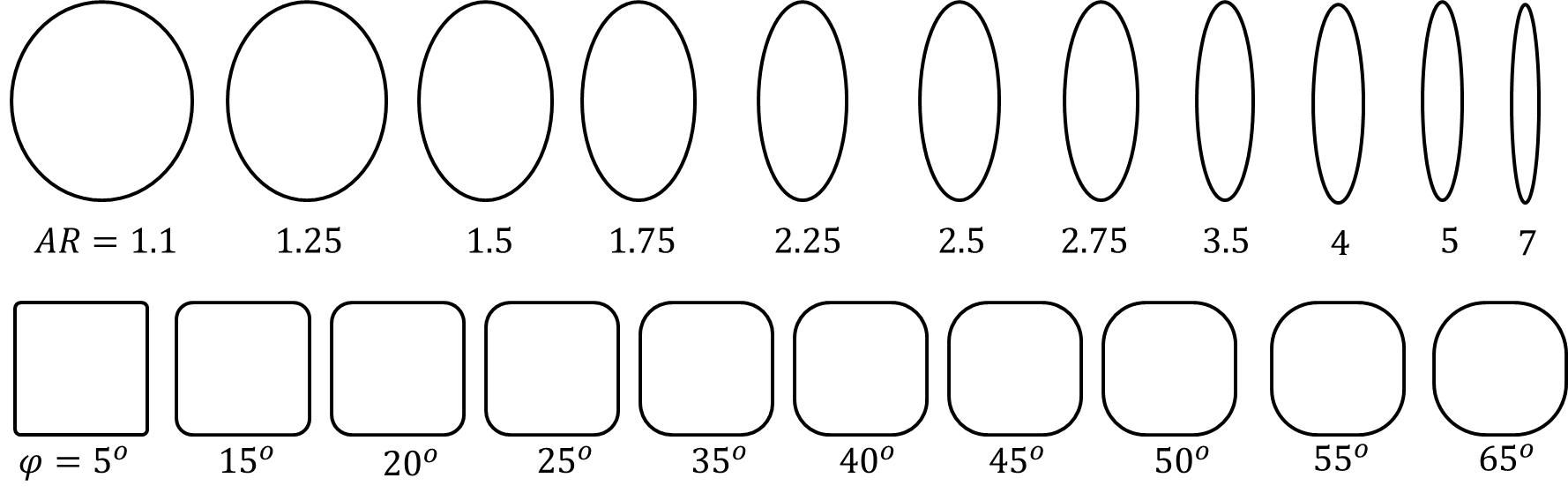}}
\caption{}
\end{subfigure}
  \caption{Problem definition and parameter space for the application of deep convolutional network: 
  (a) schematics of bluff-body geometries with relevant dimensions: aspect ratio ($AR$) of ellipses and rounding angle ($\varphi$) 
  of squares are shown, (b) known training geometry set,
  (c) vortex street and time histories of the force coefficients for a representative case: circular cylinder ($AR=1$ and $\varphi=90^o$)
  and (d) unknown geometries to be predicted by using the FOM data of the training set.}
  \label{fig:Problem}
\end{figure}
Figure \ref{fig:Problem} (a) shows the two-dimensional problem domain and the boundary conditions. A Dirichlet boundary condition $(u=U,v=0)$ is employed at the inlet and a Neumann boundary condition $(\frac{\partial u}{\partial x} =0, \frac{\partial v}{\partial x} =0, p=0)$ is applied at the outlet. The top and bottom boundaries are applied with a symmetric boundary condition $(\frac{\partial u}{\partial y} =0, v=0)$ to simulate the far-field conditions. A no-slip boundary condition is applied on the bluff body surface.
The flow domain $\Omega$ consists of a collection of non-overlapping finite elements with a body-fitted formulation with finer grid resolution in the regions of high gradients.

While Fig. \ref{fig:Problem} (a) shows the bluff body shapes 
immersed in a uniform flow speed of $U$, Figs. \ref{fig:Problem}(b) and (d) illustrate the training cases and prediction cases respectively. Even though we define two different configurations of bluff bodies, they share a common element: the circular bluff body can be considered as an ellipse with aspect ratio, $AR = 1$ and a rounded square with a rounding angle $\varphi =90^o$. Hence, we assume that the high-fidelity data of one 
configuration will be useful in predicting the wake flow characteristics of 
the other configuration. 
We perform 
the full-order simulation over a small subset of the infinite set of problem
geometries shown in Fig. \ref{fig:Problem} (b), for a laminar flow at $Re=100$ (based on the diameter $D$) to extract the velocity and pressure fields for 20 cycles of vortex shedding. Using the flow field data, we first calculate the time histories of the lift and drag coefficients and then the mean drag ($\overline{C_D}$) and the root mean square (rms) of the lift ($C_L^{rms}$) for each training case. Figure \ref{fig:Problem} (c) shows the vortex shedding pattern, the time histories of the force coefficient and the extracted time-averaged statistics for the representative case of circular shape. We feed these values as the output layer of the CNN while a distance function based on the bluff body geometry is fed as the input to the CNN. The deep neural network is trained to obtain the functional relationship between the bluff body geometry and the force dynamics such that it can predict the force  coefficients for any perturbed geometry of the training set. 
As a demonstration,  we estimate the force coefficient for all geometries 
in the set shown in  Fig. \ref{fig:Problem} (d).

\setkeys{Gin}{draft=false}
\begin{figure}[h]
\centering
\begin{subfigure}{0.5\textwidth}
\centering
\includegraphics[trim={0 0 0 0},clip,scale=0.265]{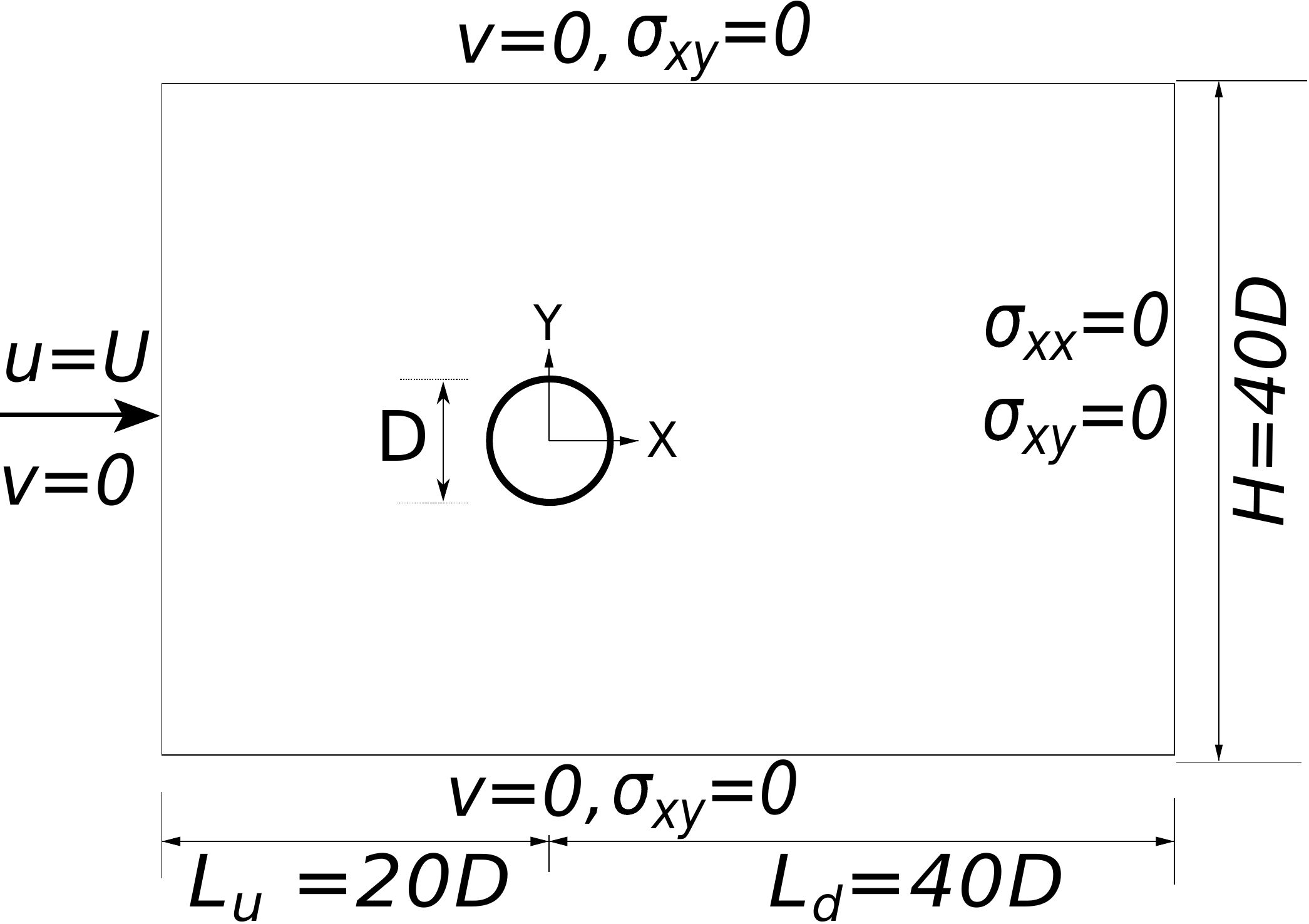}
\centering
\caption{}
\end{subfigure}~
\begin{subfigure}{0.5\textwidth}
\centering
\includegraphics[trim={0 2.5cm 0 2.5cm},clip,scale=0.245]{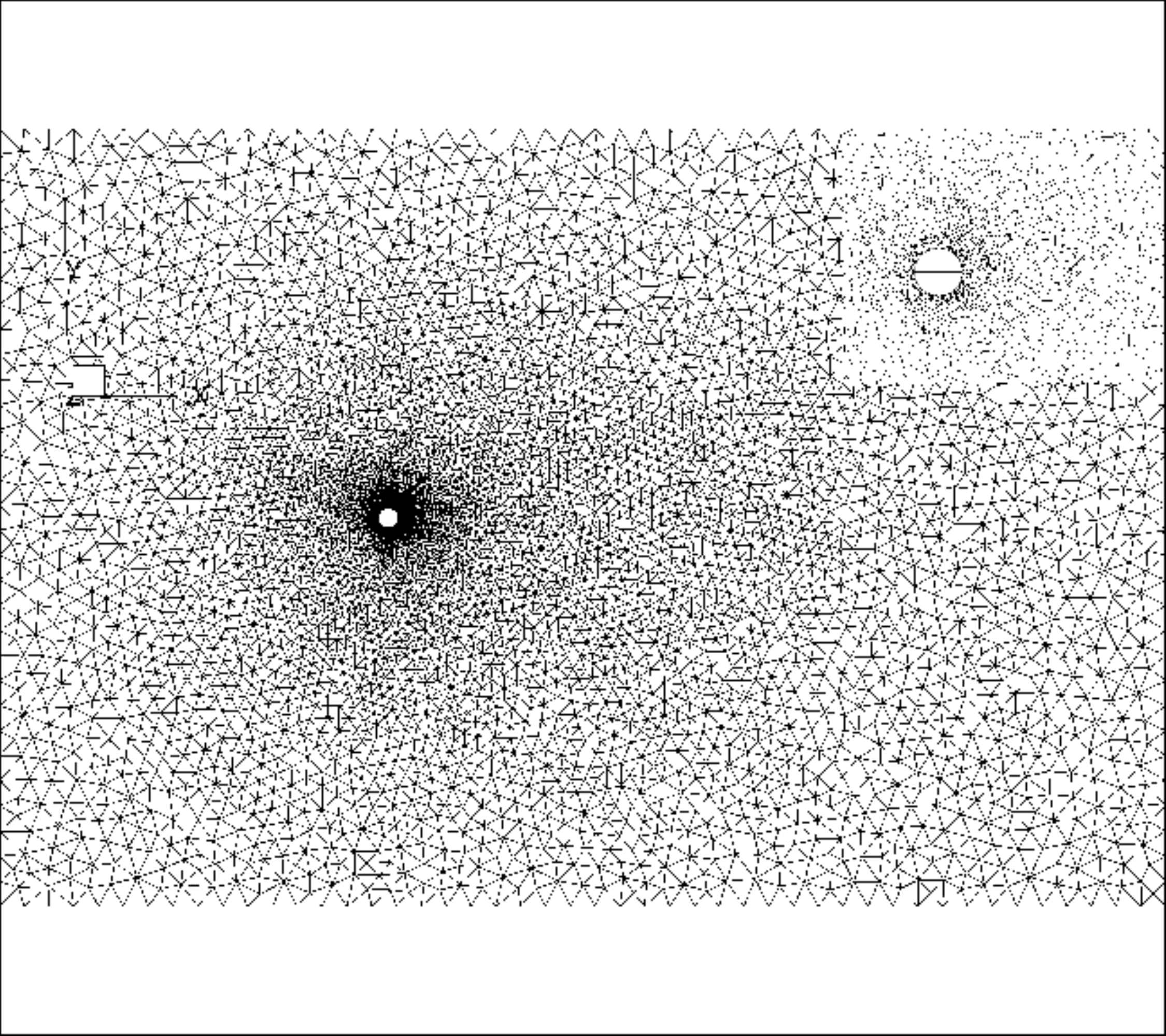}
\centering
\caption{}
\end{subfigure}
  \caption{Full order model based on Navier-Stokes equations: 
  (a) problem domain and
  (b) unstructured 2D mesh
  }
  \label{fig:Schematic}
\end{figure}
\setkeys{Gin}{draft=false}
The fluid loading is computed by integrating the surface traction considering the first layer of elements located on the cylinder surface. The instantaneous lift and drag force coefficients are defined as
\begin{align}
C_L = \frac{1}{\frac{1}{2}\rho U^2D}\int_{\Gamma}(\boldsymbol{\sigma} .\bn).\bn_y d\Gamma, \\
C_D= \frac{1}{\frac{1}{2}\rho U^2D}\int_{\Gamma}
(\boldsymbol{\sigma} .\bn).\bn_x d\Gamma,
\label{eq:CLandCD}
\end{align}
where $D$ is the cylinder dimension perpendicular to the flow direction and ${\bn_x}$ and ${\bn_y}$ are the Cartesian components of the unit normal, {$\bn$}.
The domain is discretized using an unstructured finite-element
mesh shown in Fig.~\ref{fig:Schematic}(b). There is a boundary layer mesh surrounding the bluff body and triangular mesh outside the boundary layer region. This mesh is obtained by the convergence studies conducted in \cite{jaiman_caf2016,miyanawala2016flow}. The mesh selected for the circular cylinder contains 86,638 elements. This element count slightly differs for each bluff-body geometry. Based on our previous temporal convergence studies in \cite{jaiman_caf2016,miyanawala2016flow}, we employ $\Delta t = 0.025 \left( D/U_{\infty} \right)$.

\subsection{Euclidean distance function}
We define an input function which represents the bluff body geometry and is independent of any other variable. We use the Euclidean distance from the bluff body boundary given by $\boldmath{d}(x,y)=\{d_{ij}\} \in \mathbb{R}^{m \times n}$ for a Cartesian $XY$-coordinate frame $(x\in X, y\in Y)$ centered at the bluff body center such that:
\begin{equation}
d_{ij} = \left(\min ||r(x_i,y_j) - r(x_{\Gamma},y_{\Gamma})||\right) b, \: \: i=1,2,...,m, \: \: j=1,2,...,n, 
\end{equation}
where $r$ is the distance to the domain point from the bluff-body center and $\Gamma$ represents the independent bluff body boundary points. The binary condition $b=0$ on and inside the boundary and $b=1$ otherwise.
The integers $m$ and $n$ are the matrix dimensions in the $x$ and $y$ directions, which depend on the underlying fluid domain discretization. 
In all cases, the input discrete representation of geometry is evaluated on a coarse uniform grid (i.e., $x_i,y_j$) of $0.2D$ intervals on a rectangular domain $[(-20D,-20D), (40D,20D)]$,  which
results into a 2D matrix with a size $m\times n = 201 \times 301$. The discrete interface boundary $(x_{\Gamma}, y_{\Gamma})$ represents ($x, y$) coordinates of 512 points on the bluff body boundary, which are independent of the main coarse grid. This grid resolution and the fine boundary discretization properly capture the geometry variations among different bluff bodies. 
The feed-forward process is then applied to the input geometry matrix to extract the dominant features  of the flow. As mentioned in Section \ref{sec:Feedforward} we need to tune the algorithm by optimizing a few hyper-parameters. The next section describes the hyper-parameter design performed in this particular case.

\subsection{Hyper-parameter and sensitivity analysis}
\label{HyperP}
Extreme refining and overuse of the convolution layers may cause the CNN to overfit the training dataset and make it incapable of predicting the forces for perturbed geometries of the training dataset. However, the under-utilization of convolution will increase the error of prediction. Here, we present an empirical sensitivity study 
to establish the hyper-parameter values for the best performance of 
the CNN-based learning. Specifically, we address some of the general issues  
raised by \cite{kutz2017deep} for deep learning methods in fluid flows e.g.,
the number of convolution layers, the number of filters per layer, the size of the filters and the learning rate.  

The main benchmark considered here is the maximum relative error of the drag coefficient prediction which is required to be below $5\%$ threshold for the acceptable hyper-parameter set. We evaluate this error for the training set and also select a subset of 10 target geometries ($AR = 1.5, 2.5, 4, 5, 7$ and rounding angle = $5 \degree, 20 \degree, 40 \degree, 45 \degree, 50 \degree$) which we denote as the test set. We start with the hypothesis that the CNN process with the least overall size is optimal for the predictions since it has the smallest fluid domain neighborhood representing a flow feature of dynamical importance.
Our first trial is to increase the stride which can cause our results 
to deteriorate since it increases the fluid neighborhood size of a feature 
and generates erroneous local connectivities between farther points in the fluid 
domain. For all convolution layers, we use a stride of $s_l = [1 \; \; 1]$: i.e. the convolution kernel block is moved adjacently in both directions of the 2D matrix.
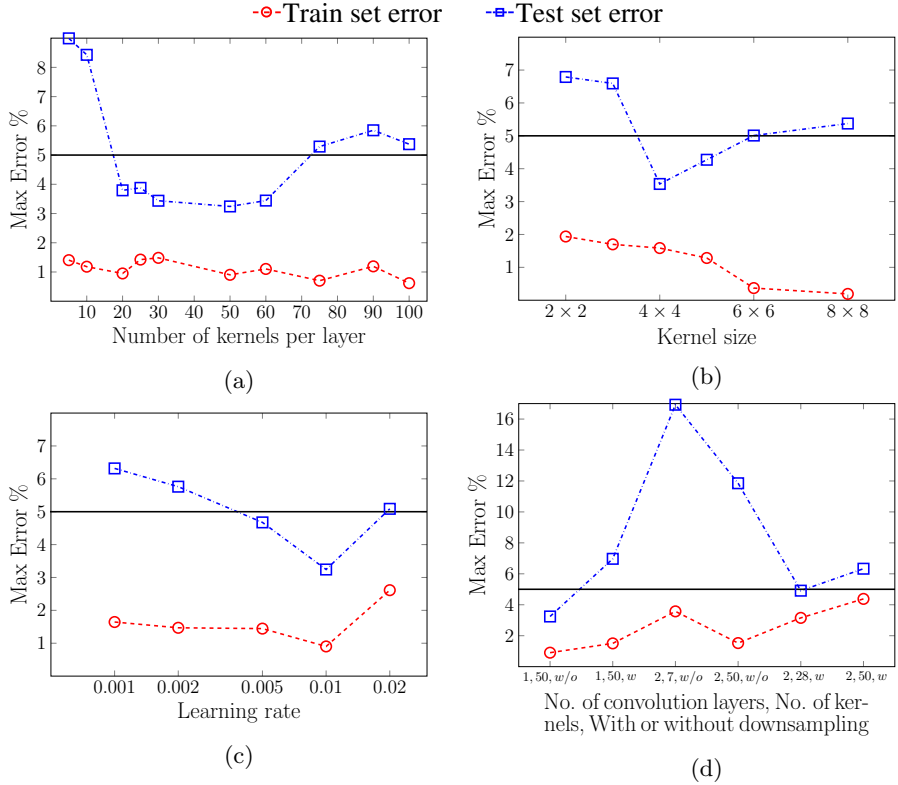
\begin{figure}[h]
\centering
\begin{subfigure}[]{\textwidth}
\centering
\includegraphics[trim={0 16cm 0 0.5cm},clip,scale=0.35]{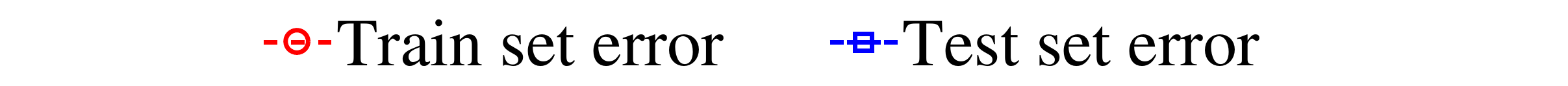}
\end{subfigure}
\pgfplotsset{every tick label/.append style={font=\Large}}
\pgfplotsset{every tick label/.append style={scale=1}}
\centering
\begin{subfigure}{0.5\textwidth}
\centering
\begin{tikzpicture}[
trim axis left, 
trim axis right, 
scale=0.5, 
baseline]
\begin{axis}[
	scale only axis,
    width=10cm,
	height=7cm,
    xlabel={\LARGE Number of kernels per layer},
    ylabel={\LARGE Max Error \%},
    xmin=0, xmax=105,
    ymin=0, ymax=9,
    xtick={10,20,30,40,50,60,70,80,90,100},
    ytick={1,2,3,4,5,6,7,8},
    yticklabel style={/pgf/number format/fixed, /pgf/number format/precision=2},
]
 
\addplot[
    mark size = 4,
    color=red,
    mark=o,
    mark options=solid,
    very thick,
    dashed,
    ]
    coordinates {
    (5,1.4048)(10,1.1812)(20,0.9495)(25,1.4236)(30,1.4834)(50,0.9029)(60,1.1036)(75,0.7023)(90,1.1923)(100,0.6172)
    };
	\addplot[
    color=blue,
    mark size = 4,
    dash dot,
    mark=square,
    mark options=solid,
    very thick,
    ]
    coordinates {
    (5,8.9926)(10,8.4292)(20,3.7936)(25,3.8775)(30,3.4375)(50,3.2423)(60,3.4428)(75,5.2904)(90,5.8480)(100,5.3725)
    };
\addplot[
    color=black,
    solid,
    very thick,
    ]
    coordinates {
    (0,5)(105,5)
    };
 \end{axis}
\end{tikzpicture}
\caption{}
\end {subfigure}~
\begin{subfigure}{0.5\textwidth}
\centering
\begin{tikzpicture}[
trim axis left, 
trim axis right, 
scale=0.5, 
baseline]
\begin{axis}[
	scale only axis,
    width=10cm,
	height=7cm,
    xlabel={\LARGE Kernel size},
    ylabel={\LARGE Max Error \%},
    xmin=1, xmax=9,
    ymin=0, ymax=8,
    xtick={2,4,6,8},
    xticklabels={$2\times 2$,$4\times4$,$6\times6$,$8\times8$},
    ytick={1,2,3,4,5,6,7},
    yticklabel style={/pgf/number format/fixed, /pgf/number format/precision=2},
]
 
\addplot[
    mark size = 4,
    color=red,
    mark=o,
    mark options=solid,
    very thick,
    dashed,
    ]
    coordinates {
    (2,1.9379)(3,1.6992)(4,1.5863)(5,1.2846)(6,0.3696)(8,0.1920)
    };
	\addplot[
    color=blue,
    mark size = 4,
    dash dot,
    mark=square,
    mark options=solid,
    very thick,
    ]
    coordinates {
    (2,6.7900)(3,6.5933)(4,3.5360)(5,4.2720)(6,5.0086)(8,5.3736)
    };
\addplot[
    color=black,
    solid,
    very thick,
    ]
    coordinates {
    (0,5)(105,5)
    };
 \end{axis}
\end{tikzpicture}
\caption{}
\end {subfigure}
\begin{subfigure}{0.5\textwidth}
\centering
\begin{tikzpicture}[
trim axis left, 
trim axis right, 
scale=0.5, 
baseline]
\begin{axis}[
	scale only axis,
    width=10cm,
	height=7cm,
    xlabel={\LARGE Learning rate},
    ylabel={\LARGE Max Error \%},
    xmin=0.0005, xmax=0.03,
    ymin=0, ymax=8,
    xmode=log,
    log ticks with fixed point,
    xtick={0.001,0.002,0.005,0.01,0.02},
    ytick={1,2,3,4,5,6,7},
    yticklabel style={/pgf/number format/fixed, /pgf/number format/precision=2},
]
 
\addplot[
    mark size = 4,
    color=red,
    mark=o,
    mark options=solid,
    very thick,
    dashed,
    ]
    coordinates {
    (0.001,1.6450)(0.002,1.4697)(0.005,1.4448)(0.01,0.9023)(0.02,2.6134)
    };
	\addplot[
    color=blue,
    mark size = 4,
    dash dot,
    mark=square,
    mark options=solid,
    very thick,
    ]
    coordinates {
    (0.001,6.3169)(0.002,5.7613)(0.005,4.6749)(0.01,3.2423)(0.02,5.0875)
    };
\addplot[
    color=black,
    solid,
    very thick,
    ]
    coordinates {
    (0.0001,5)(0.03,5)
    };
 \end{axis}
\end{tikzpicture}
\caption{}
\end {subfigure}~
\begin{subfigure}{0.5\textwidth}
\centering
\begin{tikzpicture}[
trim axis left, 
trim axis right, 
scale=0.5, 
baseline]
\begin{axis}[
	scale only axis,
    width=10cm,
	height=7cm,
    xlabel={\LARGE No. of convolution layers, No. of kernels, With or without downsampling},
    xlabel style={align=center,text width=12cm},
    ylabel={\LARGE Max Error \%},
    xmin=0.5, xmax=6.5,
    ymin=0, ymax=17,
    xtick={1,2,3,4,5,6},
    xticklabels={\normalsize $1,50,w/o$\\ \normalsize $1,50,w$\\ \normalsize $2,7,w/o$\\ \normalsize $2,50,w/o$\\ \normalsize $2,28,w$\\ \normalsize $2,50,w$\\},
    ytick={2,4,6,8,10,12,14,16},
    yticklabel style={/pgf/number format/fixed, /pgf/number format/precision=2},
]
 
\addplot[
    mark size = 4,
    color=red,
    mark=o,
    mark options=solid,
    very thick,
    dashed,
    ]
    coordinates {
    (1,0.9029)(2,1.5035)(3,3.5733)(4,1.5239)(5,3.1521)(6,4.3789)
    };
	\addplot[
    color=blue,
    mark size = 4,
    dash dot,
    mark=square,
    mark options=solid,
    very thick,
    ]
    coordinates {
    (1,3.2423)(2,6.9691)(3,16.9268)(4,11.8524)(5,4.9105)(6,6.3288)
    };
\addplot[
    color=black,
    solid,
    very thick,
    ]
    coordinates {
    (0,5)(105,5)
    };
 \end{axis}
\end{tikzpicture}
\caption{}
\end {subfigure}
  \caption{Hyper-parameter analysis: maximum error variation with (a) number of kernels, (b) kernel size, (c) learning rate and (d) number of layers. The solid lines denote the error threshold of $5\%$.}
  \label{fig:Hyperparameters}
\end{figure}
As shown in Fig. \ref{fig:Hyperparameters} (a), the least error is observed when 50 kernels are used. When a less number of kernels is used, it misses some flow features 
required to predict $\overline{C_D}$ from the bluff body geometry. 
If we increase it too much, the CNN introduces erroneous features 
increasing the prediction error. 
%
Figure \ref{fig:Hyperparameters} (b) shows that the prediction error is minimal: for the training 
set when $8 \times 8$ kernels are used and for the test set when $4 \times 4$ kernels are used. We use $4\times 4$ kernels since the priority is to predict $\overline{C_D}$ for new geometries via the generalization property of deep nets. Also, the $8 \times 8$ kernels case indicates the characteristics of a slightly over-fitting of CNN. 
The use of a smaller kernel than $4\times 4$ causes the CNN to miss the proper local connectivity as it incorporates too small neighborhoods for flow features,
larger kernels add inaccurate local connectivities to the CNN deviating 
the $\overline{C_D}$ prediction from the actual value.
At a learning rate of $0.01$, we obtain the best predictions 
as shown in Fig. \ref{fig:Hyperparameters}c. Smaller learning rates do not provide an acceptable accuracy within a prescribed number of iterations.  
If the learning rate is too large, the weights of the CNN oscillate between the same values without converging
to the minimum error.

Next, we conduct the error analysis with a different number of layers, the kernels and with and without a down-sampling layer, which is shown in Fig. \ref{fig:Hyperparameters}d. We find that the CNN performs best when a single composite layer with 50 kernels is used without the down-sampling layer. When we use more layers, it increases the fluid neighborhood size representing a local feature creating invalid local connectivities. Down-sampling drops many points of the feature map obtained by a kernel which causes a much larger fluid neighborhood to be represented by a single value. This affects adversely on the wake flow predictions where the features are highly local. 
To summarize, we use a CNN with one convolution layer consisting of 50 kernels of the size $4 \times 4$, followed by an reLU rectification layer and an SGDM back-propagation with a learning rate of 0.01. Some sample data and codes used in this work are publicly available on Github at: \url{https://github.com/TharinduMiyanawala/CNNforCFD}

\subsection{Check for overfitting}
\label{sec:Overfitting}
\begin{figure}[h]
  \centering
\pgfplotsset{every tick label/.append style={font=\Large}}
\pgfplotsset{every tick label/.append style={scale=1}}
\centering
\begin{subfigure}{0.5\textwidth}
\centering
\begin{tikzpicture}[
trim axis left, 
trim axis right, 
scale=0.5, 
baseline]
\begin{axis}[
	scale only axis,
    width=11cm,
	height=7cm,
    xlabel={\LARGE Ellipses: $AR$ \: \: \: Rounded Sq.: $\varphi$},
    xlabel style = {at={(current axis.right of origin)}, xshift=5.5ex, yshift=-3.5ex},
    ylabel={\LARGE $\overline{C_D}$},
    xmin=0, xmax=8,
    ymin=1.2, ymax=2.52,
    xtick={1,2,3,4,5,6,7},
    xticklabels={1,2,3,10,$10^o$,$30^o$,$60^o$},
    ytick={1.2,1.4,1.6,1.8,2.0,2.2,2.4},
    legend style = {draw=none},
    legend style={at={(0.6,0.8)},anchor=west},
    yticklabel style={/pgf/number format/fixed, /pgf/number format/precision=2},
]
 
\addplot[
	only marks,
    mark size = 4,
    color=black,
    mark=*,
    very thick,
    ]
    coordinates {
    (1,1.3628)(2,1.8474)(3,2.1438)(4,2.3694)(5,1.4160)(6,1.3662)(7,1.3719)
    };
     \addlegendentry{\Large FOM}; 
	\addplot[
    color=blue,
    mark size = 4,
    dashed,
    mark=square,
    mark options=solid,
    very thick,
    ]
    coordinates {
    (1,1.3662)(2,1.8657)(3,2.1122)(4,2.3867)(5,1.4024)(6,1.3873)(7,1.3592)
    };
     \addlegendentry{\Large Full-set};
     
\addplot[
    mark size = 4,
    color=green,
    mark=triangle,
    mark options=solid,
    very thick,
    solid,
    ]
    coordinates {
    (1,1.3346)(2,1.9339)(3,2.0829)(4,2.4814)(5,1.4220)(6,1.3971)(7,1.3815)
    };
     \addlegendentry{\Large 1-fold exclusion}; 
	\addplot[
    color=red,
    mark size = 4,
    dash dot,
    mark=diamond,
    mark options=solid,
    very thick,
    ]
    coordinates {
    (1,1.3640)(2,1.7848)(3,2.0960)(4,2.3684)(5,1.3950)(6,1.3675)(7,1.3493)
    };
     \addlegendentry{\Large Class-wise};
 \end{axis}
\end{tikzpicture}
\caption{}
\end {subfigure}~
\begin{subfigure}{0.5\textwidth}
\centering
\begin{tikzpicture}[
trim axis left, 
trim axis right, 
scale=0.5, 
baseline]
\begin{axis}[
	scale only axis,
    width=11cm,
	height=7cm,
    xlabel={\LARGE Ellipses: $AR$ \: \: \: Rounded Sq.: $\varphi$},
    xlabel style = {at={(current axis.right of origin)}, xshift=5.5ex, yshift=-3.5ex},
    ylabel={\LARGE Max. Error \%},
    xmin=0, xmax=8,
    ymin=0, ymax=6,
    xtick={1,2,3,4,5,6,7},
    xticklabels={1,2,3,10,$10^o$,$30^o$,$60^o$},
    ytick={1,2,3,4,5,6},
    legend style = {draw=none},
    legend style={at={(0.6,0.6)},anchor=west},
    yticklabel style={/pgf/number format/fixed, /pgf/number format/precision=2},
]
 
\addplot[
    color=black,
    mark=none,
    very thick,
    solid
    ]
    coordinates {
    (0,5)(8,5)
    };
     \addlegendentry{\Large 5\% Threshold}; 
	\addplot[
    color=blue,
    mark size = 4,
    dashed,
    mark=square,
    mark options=solid,
    very thick,
    ]
    coordinates {
    (1,0.25)(2,0.99)(3,1.47)(4,0.73)(5,0.96)(6,1.55)(7,0.92)
    };
     \addlegendentry{\Large Full-set};
     
\addplot[
    mark size = 4,
    color=green,
    mark=triangle,
    mark options=solid,
    very thick,
    solid,
    ]
    coordinates {
    (1,2.07)(2,4.68)(3,2.84)(4,4.73)(5,0.42)(6,2.26)(7,0.70)
    };
     \addlegendentry{\Large 1-fold exclusion}; 
	\addplot[
    color=red,
    mark size = 4,
    dash dot,
    mark=diamond,
    mark options=solid,
    very thick,
    ]
    coordinates {
    (1,0.09)(2,3.39)(3,2.23)(4,0.04)(5,1.48)(6,0.095)(7,1.65)
    };
     \addlegendentry{\Large Class-wise};
 \end{axis}
\end{tikzpicture}
\caption{}
\end {subfigure}
  \caption{(a) Assessment of CNN-based $\overline{C_D}$ prediction for the training set of using full-set, 1-fold exclusion and class-wise training. (b) Relative percentage error for each case.  \label{fig:Overfit}}
\end{figure}
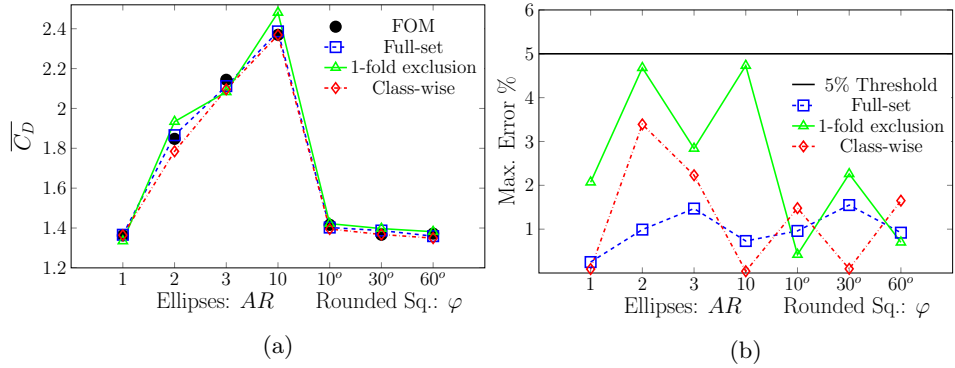

When tuning the hyper-parameters of the CNN, we face the risk of overfitting 
the algorithm to the training data. However, it is possible to check the developed CNN algorithm for overfitting without running a validation check with FOM of predictions.  
We first test the algorithm using the standard
1-fold exclusion test. Here we exclude one element at a time from the training set and predict it using the CNN trained by the other inputs. If the system is overfitting, the exclusion of one element causes the system to fail to reach the accuracy required. However, the full-set predictions will generally be more accurate than the 1-fold excluded predictions.
We also introduce a class-wise training where we 
only employ the common element: circle, one ellipse and one rounded square as the training set. If the prediction error of this test falls below the required threshold, it confirms that the algorithm is not overfitting. However, if the error of this test is higher it will not give a conclusion on overfitting. 
According to Fig. \ref{fig:Overfit}, the relative error is below the $5\%$ threshold 
for both tests which confirm that the CNN-based model does not 
overfit to the training data. 
In the next section, 
we predict the force coefficients for new bluff body geometries 
using the designed CNN.

\subsection{Force predictions using CNN-based learning}
\label{Results}
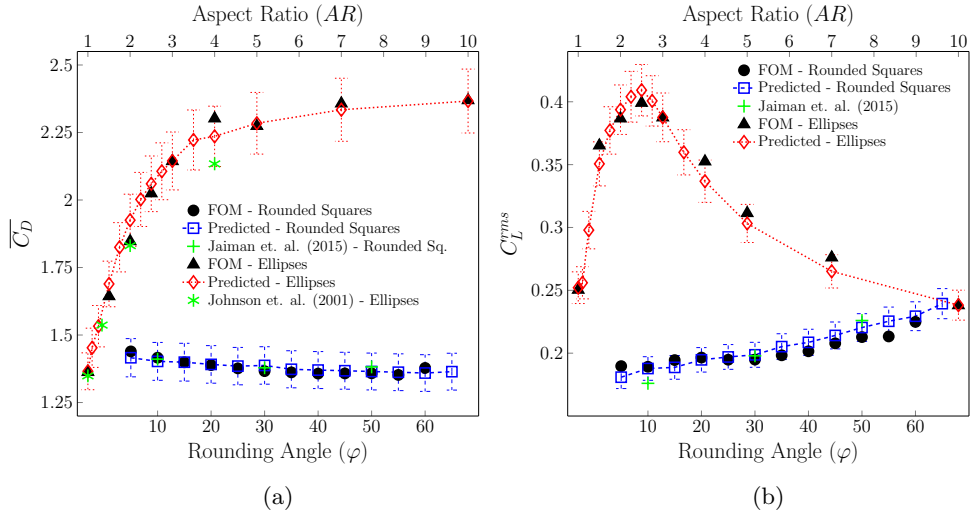
\begin{figure}
\pgfplotsset{every tick label/.append style={font=\Large}}
\pgfplotsset{every tick label/.append style={scale=1}}
\centering
\begin{subfigure}{0.5\textwidth}
\centering
\begin{tikzpicture}[
trim axis left, 
trim axis right, 
scale=0.485, 
baseline]
\begin{axis}[
	scale only axis,
    width=11cm,
	height=10cm,
    xlabel={\LARGE Rounding Angle $(\varphi)$},
    ylabel={\LARGE $\overline{C_D}$},
    xmin=-5, xmax=70,
    ymin=1.2, ymax=2.55,
    xtick={10,20,30,40,50,60},
    ytick={1.25, 1.50, 1.75,2.00,2.25,2.50},
    legend style = {draw=none},
    legend style={at={(0.4,0.45)},anchor=west},
    yticklabel style={/pgf/number format/fixed, /pgf/number format/precision=2},
    axis y line*=left,
    axis x line*=bottom
]
 
\addplot[ 
	only marks,
    mark size = 4,
    color=black,
    solid,
    mark=*,
    very thick,
    ]
    coordinates {
    (5,1.4393)(10,1.4160)(15,1.399)(20,1.3886)(25,1.3765)(30,1.3662)(35,1.3617)(40,1.3572)(45,1.3578)(50,1.3581)(55,1.3522)(60,1.3781)
    };
    \addlegendentry{\large FOM}; \label{FOMRSQ}
	\addplot[
    color=blue,
    mark size = 4,
    dashed,
    mark=square,
    mark options=solid,
    very thick,
    ]
     plot [error bars/.cd, y dir = both, y 	fixed relative = 0.05
    ]
    coordinates {
    (5,1.4159)(10,1.4024)(15,1.3994)(20,1.3912)(25,1.3848)(30,1.3873)(35,1.3731)(40,1.3702)(45,1.3677)(50,1.3649)(55,1.3622)(60,1.3592)(65,1.3643)
    };
    \addlegendentry{\large Predicted-Ellipses}; \label{PRERSQ}
    \addplot[
    only marks,
    color=green,
    mark size = 5,
    solid,
    mark=+,
    very thick,
    ]
    coordinates {
    (10,1.4105)(30,1.3786)(50,1.3818)
    };
    \addlegendentry{\large Jaiman et. al. (2015)}; \label{JaimanSq}
 \end{axis}
 \begin{axis}[
    scale only axis,
    width=11cm,
	height=10cm,
    xlabel={\LARGE Aspect Ratio $(AR)$},
    xmin=0.75, xmax=10.25,
    ymin=1.2, ymax=2.55,
    xtick={1,2,3,4,5,6,7,8,9,10},
    yticklabels={,,},
    ytick style={draw=none},
    legend style = {draw=none},
    legend style={at={(0.25,0.44)},anchor=west},
    legend cell align={left},
    yticklabel style={draw=none},
    xlabel near ticks,
    axis y line*=right,
    axis x line*=top]
    
    \addlegendimage{/pgfplots/refstyle=FOMRSQ}\addlegendentry{\large FOM - Rounded Squares}
	\addlegendimage{/pgfplots/refstyle=PRERSQ}\addlegendentry{\large Predicted - Rounded Squares}
    \addlegendimage{/pgfplots/refstyle=JaimanSq}\addlegendentry{\large Jaiman et. al. (2015) - Rounded Sq.}
    \addplot[
    only marks,
    color=black,
    mark size = 5,
    solid,
    mark=triangle*,
    very thick,
    ]
    coordinates {
    (1,1.3628)(1.5,1.6436)(2.0,1.8474)(2.5,2.0245)(3.0,2.1438)(4,2.3019)(5,2.2740)(7,2.3573)(10,2.3694)
    };
    \addlegendentry{\large FOM - Ellipses};
    
    \addplot[
    color=red,
    dotted,
    mark=diamond,
    mark size = 5,
    mark options=solid,
    very thick,
    ]
     plot [error bars/.cd, y dir = both, y 	fixed relative = 0.05
    ]
    coordinates {
    (1,1.3657)(1.1,1.4527)(1.25,1.5327)(1.5,1.6891)(1.75,1.8250)(2,1.9253)(2.25,2.0021)(2.5,2.0600)(2.75,2.1064)(3,2.1446)(3.5,2.2223)(4,2.2357)(5,2.2842) (7,2.3342)(10,2.3665)
    };
    \addlegendentry{\large Predicted - Ellipses};
    \addplot[
    only marks,
    color=green,
    mark size = 5,
    solid,
    mark=asterisk,
    very thick,
    ]
    coordinates {
    (1,1.3488)(1.3333,1.5367)(2,1.831)(4,2.1338)
    };
    \addlegendentry{\large Johnson et. al. (2001) - Ellipses};
\end{axis}
\end{tikzpicture}
\caption{}
\end {subfigure}~\hspace{3mm}
\begin{subfigure}{0.5\textwidth}
\centering
\begin{tikzpicture}[
trim axis left, 
trim axis right, 
scale=0.485, 
baseline]
\begin{axis}[
	scale only axis,
    width=11cm,
	height=10cm,
    xlabel={\LARGE Rounding Angle $(\varphi)$},
    ylabel={\LARGE $C_L^{rms}$},
    xmin=-5, xmax=70,
    ymin=0.15, ymax=0.44,
    xtick={10,20,30,40,50,60},
    ytick={0.20, 0.25,0.3,0.35,0.4},
    legend style = {draw=none},
    legend style={at={(0.5,0.85)},anchor=west},
    yticklabel style={/pgf/number format/fixed, /pgf/number format/precision=2},
    axis y line*=left,
    axis x line*=bottom
]
 
\addplot[
	only marks,
    mark size = 4,
    color=black,
    solid,
    mark=*,
    very thick,
    ]
    coordinates {
    (5,0.1897)(10,0.1892)(15,0.1945)(20,0.1965)(25,0.1948)(30,0.1949)(35,0.1983)(40,0.2015)(45,0.2078)(50,0.2128)(55,0.2133)(60,0.22486)
    };
    \addlegendentry{\large FOM}; \label{FOMRSQ}
	\addplot[
    color=blue,
    mark size = 4,
    dashed,
    mark=square,
    mark options=solid,
    very thick,
    ]
      plot [error bars/.cd, y dir = both, y 	fixed relative = 0.05
    ]
    coordinates {
    (5,0.1809)(10,0.1878)(15,0.1887)(20,0.1947)(25,0.1970)(30,0.1987)(35,0.2052)(40,0.2086)(45,0.2141)(50,0.2204)(55,0.2254)(60,0.2294)(65,0.2394)
    };
    
    \addlegendentry{\large Predicted-Ellipses}; \label{PRERSQ}
   \addplot[
    only marks,
    color=green,
    mark size = 5,
    solid,
    mark=+,
    very thick,
    ]
    coordinates {
    (10,0.1760)(30,0.1979)(50,0.2260)
    };
    \addlegendentry{\large Jaiman et. al. (2015)}; \label{JaimanSq}
 \end{axis}
 \begin{axis}[
    scale only axis,
    width=11cm,
	height=10cm,
    xlabel={\LARGE Aspect Ratio $(AR)$},
    xmin=0.75, xmax=10.25,
    ymin=0.15, ymax=0.44,
    xtick={1,2,3,4,5,6,7,8,9,10},
    yticklabels={,,},
    ytick style={draw=none},
    legend style = {draw=none},
    legend style={at={(0.4,0.85)},anchor=west},
    legend cell align={left},
    yticklabel style={draw=none},
    xlabel near ticks,
    axis y line*=right,
    axis x line*=top]
    
    \addlegendimage{/pgfplots/refstyle=FOMRSQ}\addlegendentry{\large FOM - Rounded Squares}
	\addlegendimage{/pgfplots/refstyle=PRERSQ}\addlegendentry{\large Predicted - Rounded Squares}
    \addlegendimage{/pgfplots/refstyle=JaimanSq}\addlegendentry{\large Jaiman et. al. (2015)}
    \addplot[
    only marks,
    color=black,
    mark size = 5,
    solid,
    mark=triangle*,
    very thick,
    ]
    coordinates {
    (1,0.2504)(1.5,0.3652)(2.0,0.3867)(2.5,0.3991)(3.0,0.3875)(4,0.3525)(5,0.3114)(7,0.2761)(10,0.2381)
    };
    \addlegendentry{\large FOM - Ellipses};
    
    \addplot[
    color=red,
    dotted,
    mark=diamond,
    mark size = 5,
    mark options=solid,
    very thick,
    ]
    plot [error bars/.cd, y dir = both, y 	fixed relative = 0.05
    ]
    coordinates {
    (1,0.2521)(1.1,0.2560)(1.25,0.2979)(1.5,0.3506)(1.75,0.3772)(2,0.3937)(2.25,0.4040)(2.5,0.4092)(2.75,0.4007)(3,0.3876)(3.5,0.3598)(4,0.3368)(5,0.3032)(7,0.2651)(10,0.2382)
    };
    \addlegendentry{\large Predicted - Ellipses};
\end{axis}
\end{tikzpicture}
\caption{}
\end{subfigure}
\caption{Prediction of the force coefficients: (a) mean drag and (b) rms of the lift. The results are compared with simulation results of \cite{johnson2001flow} and \cite{jaiman2015fully}.}
\label{fig:Predict}
\end{figure}

We use the convolutional neural network designed in Section \ref{HyperP} to determine the force coefficients for the perturbed geometries shown in Fig. \ref{fig:Problem} (d). We determine the error of prediction and then compare the computational cost of the full-order model and 
the CNN-based approximation.
We feed the CNN model with the new geometry input function generated for {the 
bluff body configurations in Fig. \ref{fig:Problem} (d)} and obtain the predictions of the force coefficients. 
Figure \ref{fig:Predict} compares the results of the CNN-based prediction with the 
full-order Navier-Stokes results. 
The CNN successfully predicts the force coefficients within $5\%$ error, which clearly demonstrates the generalization of our optimized convolutional nets. A direct functional relationship 
between the fluid forces and the input geometric parameters 
is established for unstable wake flow via the proposed deep learning method.

\subsection{Computational cost analysis}
\begin{table}[h]
\centering
\caption{Summary of computational resources used, speed-up and error between FOM and CNN data. While FOM is performed on a multi-core workstation and a single-core personal computer (PC) system, the CNN-based computation is done on a PC.
}{\label{FOMvsCNN}}
\begin{tabular}{l|c|c|c}
                           & FOM-HPC$^*$ & FOM-PC$^*$ &CNN$^{**}$ \\ \hline
No. of Processors          & 24      & 1       & 1 \\
Processor Speed (GHz)      & 2.60    & 2.60    & 2.60\\
RAM (GB)                   & 256     & 16      & 16\\
Elapsed time (online) (sec) & 2659.44 & 63548.21 & 69.41\\
Elapsed time(offline) (sec) & -       & -        & 287.24\\ \hline
online time per case$^{**}$ (sec) & 2659.44 & 63548.21 & 3.31\\
Total time per case$^{**}$ (sec)   & 2659.44 & 63548.21 & 16.98\\
Speed-up factor (online) &  &  & ($803^{\dagger}$, $19198^{\dagger\dagger}$) \\
Speed-up (offline + online)   &  &  & ($156^{\dagger}$, $3742^{\dagger\dagger}$) \\
Maximum Error             &  &  & $4.45\%^{\dagger}$ \\ \hline
\end{tabular}
\footnotesize{$^*$-per simulation, $^{**}$- for an input set of 21 cases, $^{\dagger}$-relative to FOM-HPC, $^{\dagger\dagger}$-relative to FOM-PC.}
\end{table}
\label{Speedup}
Table \ref{FOMvsCNN} describes the computational  resources and the elapsed time for the full-order model when running on a multicore workstation for the high-performance computing (HPC) and a single-core personal computer (PC) and the CNN-based computation  is performed on the same PC. While the parallel simulations utilize 24 processors and 256GB RAM for each case, the serial PC simulations are performed on a single processor with 16GB RAM. Note that each FOM case is a standalone and completely online computation. To obtain the force coefficients of one case, we run the FOM for 20 shedding cycles which minimize the numerical errors introduced by the initial transient response. Hence every simulation has to run until $\sim tU/D = 250 $ or $\sim 10000$ of $\Delta t U/D = 0.025$ time steps. 
The FOM performs the online calculations described in Section \ref{Sec:MatrixFormFOM} for each case. For a mesh consisting of $M$ elements, the GMRES method performs $\sim (2M^2 + NM)$ floating-point operations for the matrix-vector products and additions in the $N^{th}$ iteration for the $N^{th}$ Krylov subspace vector. Hence, for a single time step, it requires $\sim(2M^2+N^2M)$ operations. Since $N \ll M$, the total number of operations performed in the FOM solver for $T$ time steps is $\sim O(M^2T)$. In this study, $M\sim O(10^5)$ and $T\sim O(10^4)$, hence the number of operations is $\sim O(10^{14})$ per case. 
Unlike the FOM computation via finite-element GMRES approach, the CNN-based deep learning prediction has an offline training process and then the force predictions are obtained for the required inputs. This offline process gives a library of kernels and weights which are used for all predictions where in the case of full-order simulations the high-fidelity data of one FOM simulation is generally not utilized for the next FOM simulation. 
\setkeys{Gin}{draft=false}
\begin{figure}[h]
  \centering
\includegraphics[trim={0 0 0 0},clip,scale=0.032]{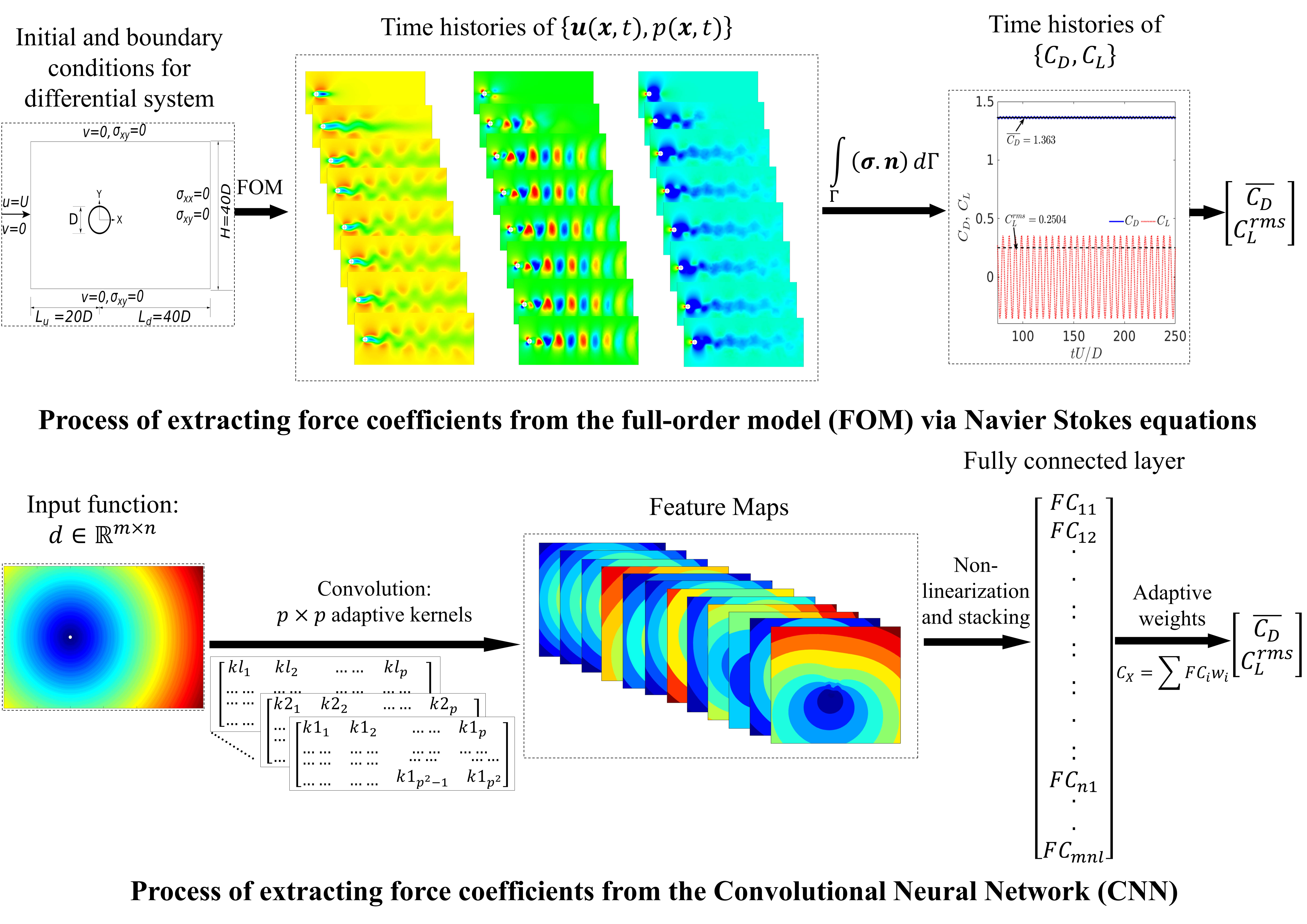}
  \caption{Comparison of the processes of extracting the force coefficients from the FOM and CNN for the flow past a bluff body. Here $\sigma (\bf{u},p)$ is the Cauchy stress tensor and $C_X=[\overline{C_D},C_{L}^{rms}]$. The convolution layer has $`l'$ layers with size $p\times p$ where $kr_i$ is the $i^{th}$ component of the $r^{th}$ layer. $FC_i$ is the $i^{th}$ element of the fully connected layer.   \label{fig:FOM_CNN}}
\end{figure}
\setkeys{Gin}{draft=false}

Figure \ref{fig:FOM_CNN} illustrates a comparison of the online processes of FOM and CNN. The online prediction phase of CNN is an extremely fast process since it requires far less number of floating point operations than FOM. For a CNN consisting of an input matrix with $M$ elements and $l$ number of $p\times p$ kernels, the convolution process requires only $\sim 2Mp^2l$ operations. The non-linearization and the fully-connected layer mapping require further $\sim 3Ml$ operations which estimate the total number of operations as $\sim Ml(2p^2+3)$ for a CNN consists of a single convolution, rectifier, and fully-connected layer.  

In this study $M\sim O(10^5)$, $l=50$ and $p=4$ which assesses the number of operations per case at $\sim O(10^8)$. Theoretically, the online process of CNN is $O(10^6)$ times faster than the FOM.
Due to the high efficiency of the CNN-based deep learning, the force predictions for all 21 cases are computed at once. Using the serial computation power of a PC, it gives $\sim 800$ speed up in comparison to the FOM simulations on an HPC which has 24 times the processing power and 16 times the RAM in contrast to the single-processor PC. Even with the offline training process, the CNN-based prediction is $\sim 150$ times faster than the HPC-FOM. When the CNN-based deep learning computation and the FOM simulation are performed on the same PC, the CNN is $\sim 19000$ times in the online phase and $\sim 3700$ times in the whole process faster than the FOM analysis. The maximum relative error observed in the CNN predictions is below than 5\% threshold, which is sufficient for a practical purpose. Another important aspect of CNN-based deep learning is that increasing the number of inputs does not increase the prediction time significantly. Hence a sufficiently large number of predictions can be obtained efficiently, i.e. in $\sim O(10)$ seconds.

In summary, the CNN-based deep learning model predicts the force coefficients within $5\%$ accuracy at $\sim O(10^4)$ speed-up per single new geometry using a small fraction of the 
computational resources as compared to its FOM counterpart. This allows the maximum utilization of the high-fidelity data obtained from the full-order methods. 
The CNN-based prediction 
has a very attractive speed versus accuracy trade-off for the design optimization process and the feedback control of bluff body flows. 
Furthermore, the CNN significantly reduces the computational resource requirement 
enabling the design space exploration of wake 
flow problems on personal computers and devices.

\section{Discussion}
Through our CNN-based deep learning results, we are successfully able to identify  the variation of the force coefficients with the two representative geometric perturbations: the aspect ratio of an ellipse and the rounding angle of a rounded square. In the present study, Eq.~(\ref{eq:CLandCD}) links the instantaneous force coefficients with the pressure $(p)$ and velocity $(\boldsymbol{u})$ fields, which are governed by the Navier-Stokes equations as functions of physical and geometric parameters e.g., $Re$ and cylinder geometry. 
In a broader sense, the full-order model provides a functional map between the geometric perturbations and the force coefficients, i.e. for the case of ellipses:
\begin{align}
\overline{C_D}(\xx) = f(AR(\xx)) \\
C_L^{rms}(\xx) = g(AR(\xx)).
\end{align}
The functions $f$ and $g$ comprise the force contributions from the flow features such as the shear layer, the near-wake, the vortex street, the time and space discretizations, the surface integration of Eq. (\ref{eq:CLandCD}) and the time averaging of the instantaneous forces. Each of the input-output pair generated by the FOM analysis is independent, contains a large amount of high-fidelity data and consumes significant computational effort. While the FOM provides a great insight into the flow physics, it is generally not suitable for the iterative optimization and the efficient engineering design study. We have shown that the CNN-based deep kernel learning provides a functional relationship between the flow parameter and design parameter and it can also capture dynamically important patterns.
In this section, we further elaborate 
our results on the force prediction for varying aspect ratios.
In particular, the variation of the lift coefficient with the aspect ratio $(AR)$ of ellipses (Fig. \ref{fig:Predict} (b)) shows an interesting pattern. It has a maximum in the vicinity of $AR=2.5$. Here we briefly examine the reasons for this behavior and investigate the capability of CNN-based deep learning to capture it.

\subsection{Interpretation of flow features influencing lift force}
\begin{figure}
  \centering
\pgfplotsset{every tick label/.append style={font=\Large}}
\pgfplotsset{every tick label/.append style={scale=1}}
\centering
\begin{minipage}[c][15cm][t]{0.5\textwidth}
\vspace*{\fill}
 \centering
\includegraphics[trim={0 0 0 0},clip,scale=0.09]{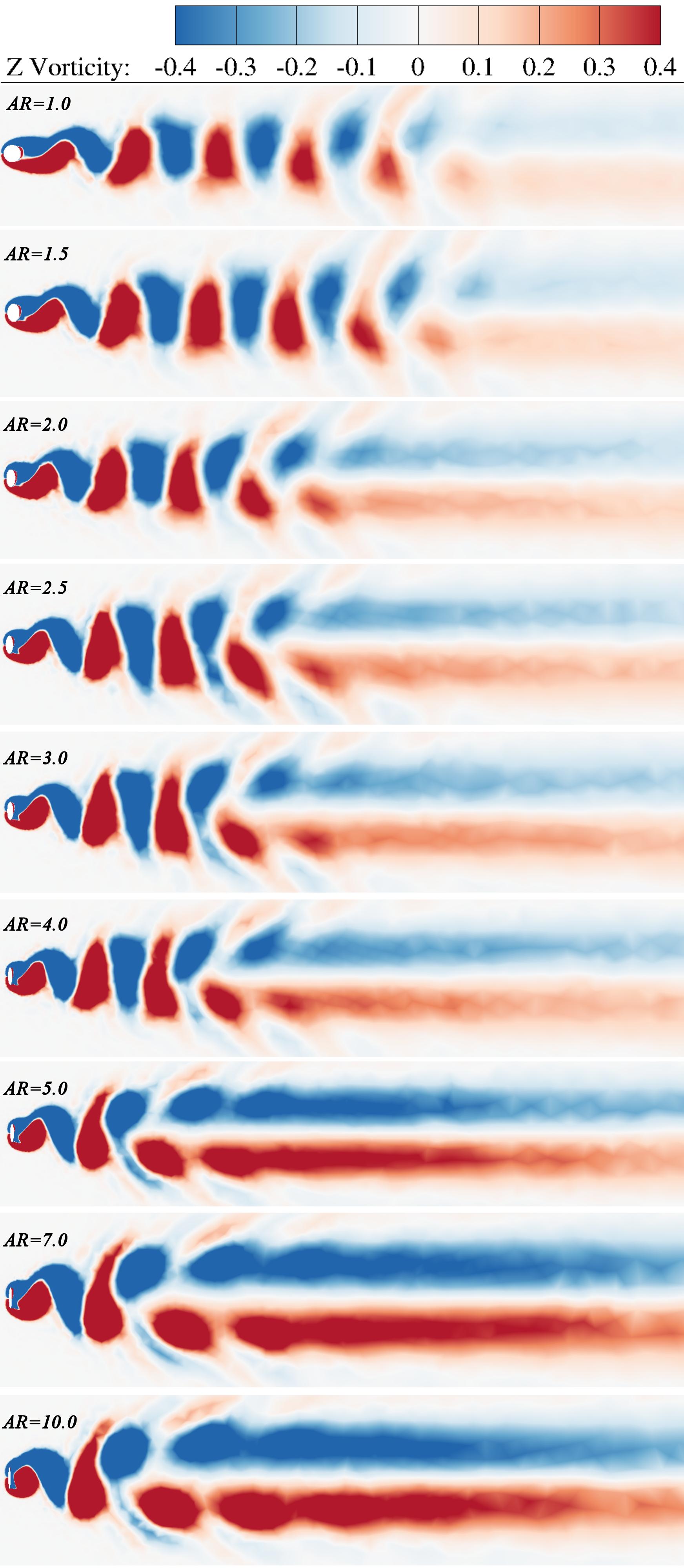}
\subcaption{}
\end{minipage}%
\begin{minipage}[c][15cm][t]{0.5\textwidth}
\vspace*{\fill}
 \centering
\includegraphics[trim={0 0 0 0},clip,scale=0.35]{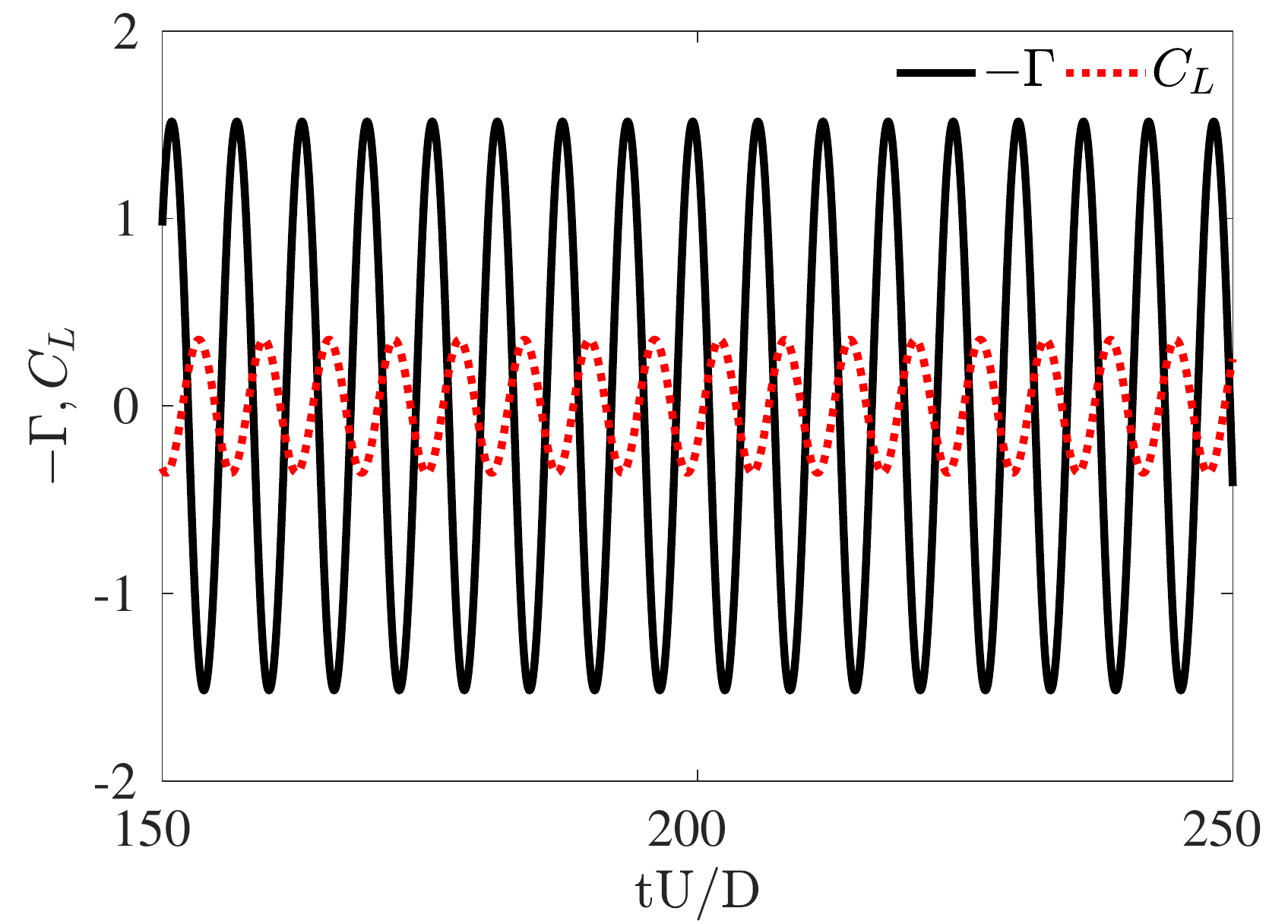}
\subcaption{}
\begin{tikzpicture}[
trim axis left, 
trim axis right, 
scale=0.44, 
baseline]
\begin{axis}[
    scale only axis,
    width=11cm,
	height=7cm,
    xlabel={\LARGE Aspect Ratio $(AR)$},
    ylabel={\LARGE $\Gamma^{rms}$},
    xmin=0.75, xmax=10.25,
    ymin=1.0, ymax=1.9,
    xtick={1,2,3,4,5,6,7,8,9,10},
    ytick={1.0,1.2,1.4,1.6,1.8},
    legend style = {draw=none},
    legend style={at={(0.3,0.6)},anchor=west},
    legend cell align={left},
    yticklabel style={draw=none},
    xlabel near ticks,
    axis y line*=left,
    axis x line*=bottom
]
\addplot[
    mark size = 4,
    color=black,
    solid,
    mark=*,
    very thick,
    solid,
    ]
    coordinates {
    (1,1.0964)(1.5,1.4166)(2.0,1.5715)(2.5,1.7200)(3.0,1.7723)(4,1.8248)(5,1.8022)(7,1.7570)(10,1.5855)
    };
    \addlegendentry{\large $\Gamma^{rms}$}; \label{Gamma}
    
 \end{axis}
 \begin{axis}[
    scale only axis,
    width=11cm,
	height=7cm,
    ylabel={\LARGE $L_r$},
    xmin=0.75, xmax=10.25,
    ymin=0.8, ymax=3.1,
    xticklabels={,,},
    ytick style={draw=none},
    legend style = {draw=none},
    legend style={at={(0.3,0.6)},anchor=west},
    legend cell align={left},
    axis y line*=right,
    axis x line*=top]
    
    \addlegendimage{/pgfplots/refstyle=Gamma}\addlegendentry{\large Near wake circulation $(\Gamma^{rms})$}
    \addplot[
    color=red,
    dotted,
    mark=diamond*,
    mark size = 5,
    mark options=solid,
    very thick,
    ]
    coordinates {
    (1,2.9787)(1.5,2.2340)(2.0,1.9713)(2.5,1.7021)(3.0,1.5957)(4,1.4894)(5,1.4128)(7,1.2830)(10,0.9149)
    };
    \addlegendentry{\large Recirculation length $(L_r)$};
    
\end{axis}
\end{tikzpicture}\\
\subcaption{}
\end{minipage}
  \caption{Analysis of flow past over elliptical cylinders of varying aspect ratios. Flow features affecting the lift variation with aspect ratio: (a) Development of vortex shedding patterns observed for different aspect ratios, (b) Time traces of near wake circulation and lift coefficient for $AR=1.0$, and (c) Variation of near wake circulation and recirculation length with $AR$. Note that $\Gamma$ and $L_r$ values presented here are non-dimensionalized by $UD$ and $D$, respectively.   \label{fig:AReffect}}
\end{figure}
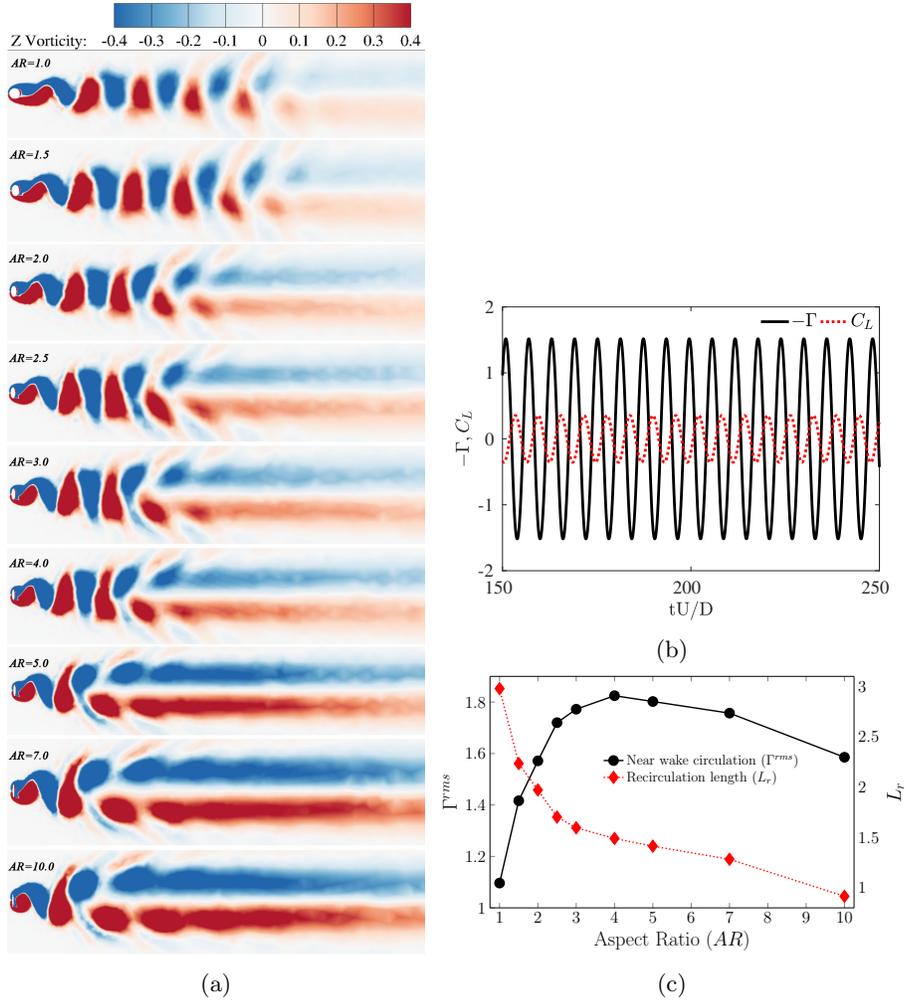

The lift force of a bluff body is affected by many flow features. Here we discuss two such features namely, the near wake circulation and the recirculation length, and their variations with the aspect ratio of the elliptical bluff body.
According to the Kutta-Joukowski's theorem for an inviscid flow around a streamline body that the lift force per unit span can be expressed as: $F_L = -\rho^\mathrm{f} U \Gamma$,
where $\Gamma$ is the circulation given by the line integral $\oint_S \boldsymbol{u}^\mathrm{f}\cdot \boldsymbol{dS}$ evaluated on a closed contour $S$ enclosing the streamlined body. Although the viscous flow past a bluff body does not satisfy some of the conditions, it suggests the important relationship: the lift force has some proportional relationship with the circulation of the vortical wake. Here, we observe the vortex patterns and calculate the total circulation of the wake to describe the lift variation observed due to the perturbation of the ellipse geometry.
Figure \ref{fig:AReffect} (a) shows the vorticity patterns of the wake of different ellipses. When the aspect ratio is changed from the circular cylinder $(AR=1.0)$ to $AR=10.0$, the vortex pattern changes from a standard Karman street to a two-layered vortical wake pattern. This behavior is also observed for $Re=100$ in the recent study by \cite{thompson2014low}. It is clear that the vortex intensity and the recirculation length differs according to the aspect ratio of an elliptical bluff body. We quantify the total circulation of the near wake by considering the closed rectangular contour bounded by $(-1,-1.4)D, (3.5,-1.4)D, (3.5,1.4)D$ and $(-1,1.4)D$. Apart from the $x=3.5D$ line segment, all the other line segments pass through irrotational regions of the flow. The circulation is taken positive in the anti-clockwise sense. Figure \ref{fig:AReffect} (b) shows the time traces of  the near wake circulation and the lift coefficient. As expected, the Circulation and the lift signals have the same frequency as the Strouhal frequency, but with a phase difference.
Unlike for a streamlined body, the lift force on a bluff body also depends on the distance to the vortices in the wake. We present the circulation and the recirculation length variation with respect to the aspect ratio in Fig. \ref{fig:AReffect} (c). Note that the recirculation length is calculated as the distance to the closest vortex core just after the end of its shedding from the bluff body. The near wake circulation has a maximum of $AR=4.0$ and the recirculation length monotonically decrease with $AR$.

The predictions based on our CNN-based deep learning rely on extracting feature patterns as described here. Due to this, unlike the FOM analysis, it utilizes the full-order results of all input simulations to determine the required flow dynamic property. We further discuss the interpretation of the learned kernels in terms of the feature extractions.

\subsection{Convolution kernels}
\begin{figure}[h]
  \centering
\includegraphics[trim={3cm 1cm 2cm 1cm},clip,scale=0.4]{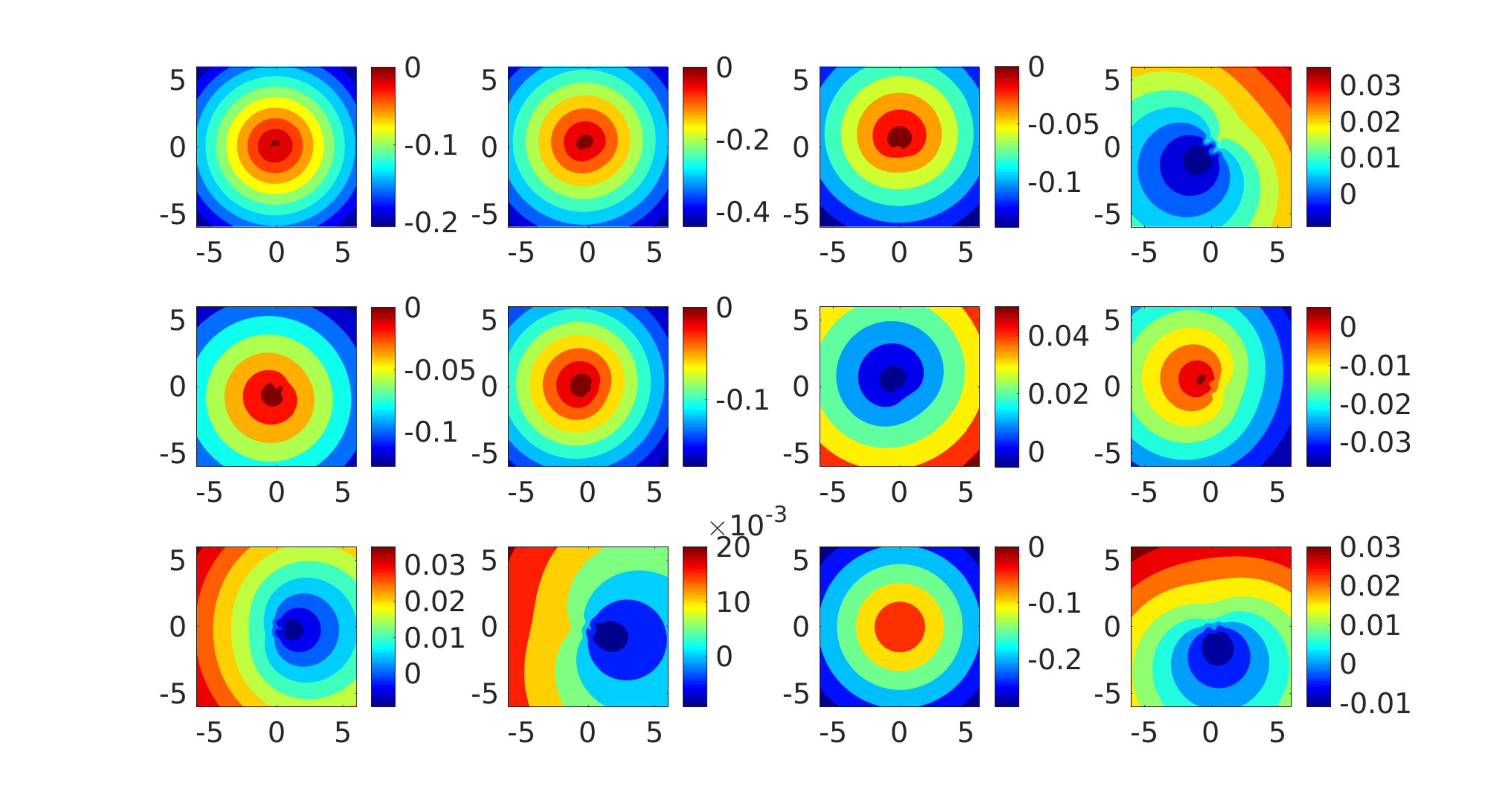}
  \caption{Feature maps obtained by critical filters of the convolutional neural network for the flow past a circular cylinder. The maps are presented in the near cylinder $5D\times 5D$ fluid domain. Note that the scales are not uniform due to the different kernel weight ranges retrieved by the adaptive training process. \label{fig:Filters}}
\end{figure}
We have demonstrated that the CNN process is efficient and accurate for the prediction of unsteady forces acting on a bluff body due to the vortex shedding process. The success of CNN-based deep learning lies in the reasonable capturing of flow features affecting the forces on the bluff body through the learning process. The main learning operation occurs at the adaptation of convolution kernels using the stochastic gradient descent method.
The learned discrete kernel is a small matrix of numbers which makes it hard to visualize and interpret. However, by applying convolution operation on the input matrix using individual kernels separately, this will produce the individual feature map of each kernel.
For example, Fig. \ref{fig:Filters} shows some of the critical feature maps extracted at the convolution layer of the circular cylinder.
These feature maps provide the encoding of features and their locations for the convolutional layers.
While the location of a feature explicitly represents where it is, the response of a feature (i.e. what it is) is encoded implicitly. 
Carefully designed deep networks based on the convolutional feature maps can extract the flow dynamical features.
It is worth noting that the feature maps will not necessarily display commonly seen flow features such as the shear layer and the Karman vortex street. In most cases, these features are reconstructed by a nonlinear combination of many kernels. Hence, the learning process is not limited to the learning of kernels but also includes learning the proper combination of kernels for the prediction process.
Next, we investigate whether the CNN-based learning is capable of predicting the lift force variation generated by the FOM even it is fed by different input cases.

\subsection{Generality of feature extraction capability}
\begin{figure}[h]
\pgfplotsset{every tick label/.append style={font=\LARGE}}
\pgfplotsset{every tick label/.append style={scale=1}}
\centering
\begin{tikzpicture}[spy using outlines={rectangle, magnification=2.5,connect spies},
trim axis left, 
trim axis right, 
scale=0.5, 
baseline]

\begin{axis}[xshift=16cm, yshift=5.5cm,
    width=5cm,
	height=5cm,
    ylabel={\large Max. error \%},
    xmin=0.5, xmax=6.5,
    ymin=2, ymax=75,
    ymode=log,
    log ticks with fixed point,
    xtick={1,2,3,4,5,6},
    ytick={1,5,10,20,40},
    ticklabel style = {font=\large},
	]
    \addplot[
    color=black,
    solid,
    very thick,
    ]
    coordinates {
    (0,5)(6,5)
    };
    \addplot[
    color=red,
    dotted,
    mark=square,
    mark size = 3,
    mark options=solid,
    very thick,
    ]
    coordinates {
    (1,44.01)
    };
    \addplot[
    color=blue,
    dashed,
    mark=triangle,
    mark size = 3,
    mark options=solid,
    very thick,
    ]
    coordinates {
    (3,4.27)
    };    
    \addplot[
    color=green,
    dash dot,
    mark=diamond,
    mark size = 3,
    mark options=solid,
    very thick,
    ]
    coordinates {
    (4,10.16)
    };
    \addplot[
    color=magenta,
    dash dot dot,
    mark=asterisk,
    mark size = 3,
    mark options=solid,
    very thick,
    ]
    coordinates {
    (2,13.05)
    };
    \addplot[
    color=cyan,
    solid,
    mark=o,
    mark size = 3,
    mark options=solid,
    very thick,
    ]
    coordinates {
    (5,9.34)
    };
    \addplot[
    color=brown!60!black,
    densely dashed,
    mark=+,
    mark size = 5,
    mark options=solid,
    very thick,
    ]
    coordinates {
    (6,4.4539)
    };
\end{axis}

 \begin{axis}[
    width=22cm,
 	height=16cm,
    xlabel={\LARGE Aspect Ratio $(AR)$},
    ylabel={\LARGE $C_L^{rms}$},
    xmin=0.75, xmax=10.25,
    ymin=0.20, ymax=0.42,
    xtick={1,2,3,4,5,6,7,8,9,10},
    ytick={0.20,0.22,0.24,0.26,0.28,0.30,0.32,0.34,0.36,
    0.38,0.40,0.42},
    legend style = {draw=none},
    legend style={at={(0.15,0.2)},anchor=west},
    legend cell align={left},
    legend style={fill=none},
	]
    
    \addplot[
    only marks,
    color=black,
    mark size = 4,
    solid,
    mark=*,
    very thick,
    ]
    plot [error bars/.cd, y dir = both, y 	fixed relative = 0.05
    ]
    coordinates {
    (1,0.2504)(1.5,0.3652)(2.0,0.3867)(2.5,0.3991)(3.0,0.3875)(4,0.3525)(5,0.3114)(7,0.2761)(10,0.2381)
    };
    \addlegendentry{\LARGE FOM};
    
    \addplot[
    color=red,
    dotted,
    mark=square,
    mark size = 5,
    mark options=solid,
    very thick,
    ]
    coordinates {
    (1,0.2502)
    (1.5,0.3195)
    (2,0.3867)
    (2.5,0.3926)
    (3,0.3872)
    (4,0.3747)(5,0.3651)(7,0.3524)(10,0.3429)
    };
    \addlegendentry{\LARGE $AR=\{1, 2, 3\}$};

    \addplot[
    color=magenta,
    dash dot dot,
    mark=asterisk,
    mark size = 5,
    mark options=solid,
    very thick,
    ]
    coordinates {
    (1,0.2564)
    (1.5,0.3367)
    (2,0.3868)
    (2.5,0.3817)
    (3,0.3495)
    (4,0.3065)(5,0.2816)(7,0.2560)(10,0.2393)
    };
    \addlegendentry{\LARGE $AR=\{1, 2, 10\}$};
    
    \addplot[
    color=blue,
    dashed,
    mark=triangle,
    mark size = 5,
    mark options=solid,
    very thick,
    ]
    coordinates {
    (1,0.2611)
    (1.5,0.3663)
    (2,0.3888)
    (2.5,0.3950)
    (3,0.3845)
    (4,0.3391)(5,0.3020)(7,0.2645)(10,0.2377)
    };
    \addlegendentry{\LARGE $AR=\{1, 3, 10\}$};
    
    \addplot[
    color=green,
    dash dot,
    mark=diamond,
    mark size = 5,
    mark options=solid,
    very thick,
    ]
    coordinates {
    (1,0.2666)
    (1.5,0.3648)
    (2,0.3909)
    (2.5,0.3810)
    (3,0.3581)
    (4,0.3257)(5,0.3049)(7,0.2805)(10,0.2623)
    };
    \addlegendentry{\LARGE $AR=\{1, 4, 10\}$};
    
    \addplot[
    color=cyan,
    solid,
    mark=o,
    mark size = 5,
    mark options=solid,
    very thick,
    ]
    coordinates {
    (1,0.2492)
    (1.5,0.3311)
    (2,0.3808)
    (2.5,0.3996)
    (3,0.3865)
    (4,0.3459)(5,0.3095)(7,0.2672)(10,0.2372)
    };
    \addlegendentry{\LARGE $AR=\{1, 3, 10\}^{\dagger}$};
    \addplot[
    color=brown!60!black,
    densely dashed,
    mark=x,
    mark size = 5,
    mark options=solid,
    very thick,
    ]
    coordinates {
    (1,0.2521)
    (1.5,0.3506)
    (2,0.3937)
    (2.5,0.4092)
    (3,0.3876)
    (4,0.3368)(5,0.3032)(7,0.2651)(10,0.2382)
    };
    \addlegendentry{\LARGE $AR=\{1, 2, 3, 10\}$};
\end{axis}
\end{tikzpicture}
\caption{Lift force prediction using different training sets. The legend states the cases used as the training set. ${\dagger}$ denotes that only ellipses are used for training. All other cases are trained together with the set of rounded square cases: $\varphi=\{10^o, 30^o, 40^o, 60^o\}$. Inset shows 
the maximum error of each prediction.
}
\label{fig:InputSelect}
\end{figure}
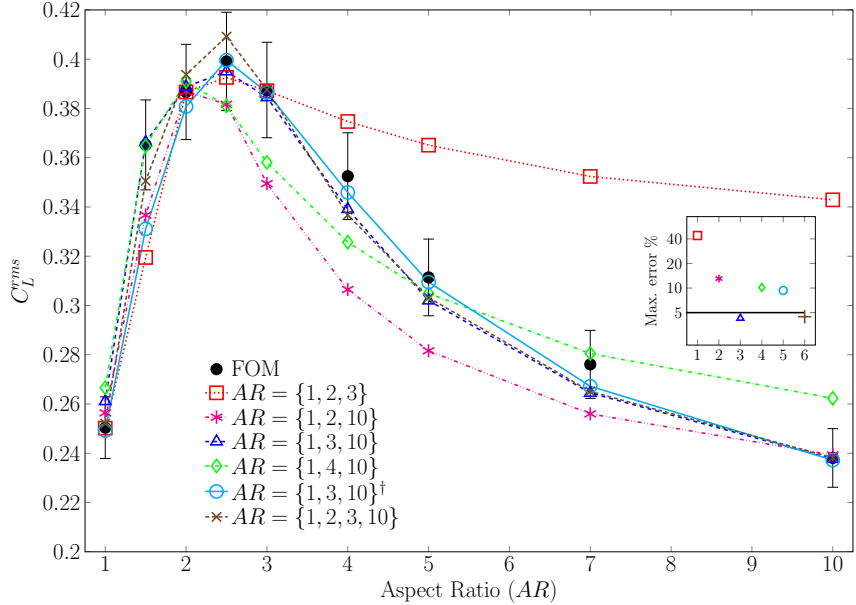
The predictions presented in Fig. \ref{Results} are generated by a small subset of input-output combinations obtained from the full-order Navier-Stokes simulations. However, the CNN-based deep learning does not operate as a traditional curve-fitting tool whereby the single-valued input-output combinations are fitted to match a predetermined general curve formula. The input to the CNN-based deep learning is not a single value but a matrix of function values derived from the single value input. This makes the CNN capable of extracting the flow features and eventually much more data efficient than standard curve-fitting and regression tools. 

\begin{figure}[tb]
\pgfplotsset{every tick label/.append style={font=\LARGE}}
\pgfplotsset{every tick label/.append style={scale=1}}
\begin{center}
\begin{tikzpicture}[
scale=0.575, 
]
 \begin{axis}[
    width=20cm,
	height=15cm,
    xlabel={\LARGE Aspect Ratio $(AR)$},
    ylabel={\LARGE $\overline{C_D}$},
    xmin=0.75, xmax=10.25,
    ymin=1.2, ymax=2.55,
    xtick={1,2,3,4,5,6,7,8,9,10},
    ytick={1.2,1.4,1.6,1.8,2.0,2.2,2.4},
    legend pos=south east,
    legend style = {draw=none},
    legend cell align={left},
    yticklabel style={draw=none},
    ylabel near ticks,
    ]
    
    \addplot[
    only marks,
    color=black,
    mark size = 5,
    solid,
    mark=*,
    very thick,
    ]
    plot [error bars/.cd, y dir = both, y 	fixed relative = 0.05
    ]
    coordinates {
    (1,1.3628)(1.5,1.6436)(2.0,1.8474)(2.5,2.0245)(3.0,2.1438)(4,2.3019)(5,2.2740)(7,2.3573)(10,2.3694)
    };
    
    \addplot[
    color=blue,
    dotted,
	mark=none,
    mark size = 5,
    mark options=solid,
    very thick,
    ]
    coordinates {
    (1,1.3657)(1.1,1.4527)(1.25,1.5327)(1.5,1.6891)(1.75,1.8250)(2,1.9253)(2.25,2.0021)(2.5,2.0600)(2.75,2.1064)(3,2.1446)(3.5,2.2223)(4,2.2357)(5,2.2842) (7,2.3342)(10,2.3665)
    };
    
    \addplot[
    color=red,
    dash dot,
	mark=none,
    mark size = 5,
    mark options=solid,
    very thick,
    ]
    coordinates {
    (1,1.3601031)(1.1,1.4448011)(1.25,1.5413543)(1.5,1.6703275)(1.75,1.7627059)(2,1.8622334)(2.25,1.9468905)(2.5,2.0175846)(2.75,2.0729237)(3,2.1185319)(3.5,2.1860948)(4,2.2344377)(5,2.2994306) (7,2.3683755)(10,2.4140677)
    };
    
    
    \addplot[
    color=green,
    dashed,
	mark=none,
    mark size = 5,
    mark options=solid,
    very thick,
    ]
    coordinates {
    (1,1.3072598)(1.1,1.4171052)(1.25,1.5498396)(1.5,1.7130691)(1.75,1.8302459)(2,1.9185460)(2.25,1.9860960)(2.5,2.0401628)(2.75,2.0842361)(3,2.1211967)(3.5,2.1769161)(4,2.2180035)(5,2.2740664) (7,2.3339183)(10,2.3751998)
    };
    
    
    
    \legend{\LARGE FOM, \LARGE CNN - Interpolation, \LARGE CNN - Extrapolation set 1, \LARGE CNN - Extrapolation set 2}
\end{axis}
\end{tikzpicture}
\caption{\color{red} Prediction of mean drag as an extrapolation problem:- extrapolation set 1: $AR = 1,2,3,7$ and $\phi=10^0, 30^0, 60^0$, and extrapolation set 2: $AR = 2,3,5,7$ and $\phi=10^0, 30^0, 60^0$.}
\label{fig:ExpolCNN}
\end{center}
\end{figure}
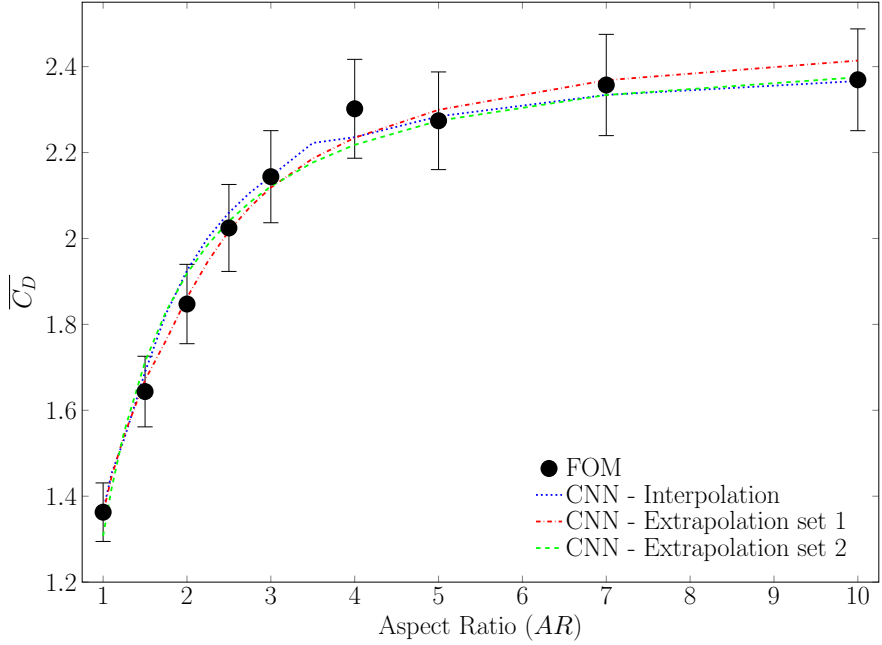

Figure \ref{fig:InputSelect} shows the performance of CNN-based learning when fed with different combinations of inputs. We find that the predictions perform best when the extremes of the geometric variations (e.g., in the case of ellipses: $AR = 1$ and $10$) are included.
Furthermore, when we include $AR=3$ case, all the predictions are accurate within 5\% error margin of the FOM results. The performance of CNN-based deep learning becomes slightly better when $AR=2$ case is also added to the training set. The interesting observation is that the predictions are significantly improved when four cases from the rounding angle problem $(\varphi = 10^o, 30^o, 40^o$ and $60^o)$ are included in the training process. This emphasizes another advantage of CNN-based prediction that we may use the data of a similar problem to enhance the accuracy of the data-driven predictions. The most significant result is that the CNN has accurately captured the maximum lift in the vicinity of $AR=2.5$ even with just three input-output pairs, which clearly demonstrates the high data efficiency of the CNN-based prediction process. It also suggests that the CNN-based learning has accurately captured the flow features in predicting the force coefficients.
We observe that when the analysis is considered as an interpolation problem, we can obtain the most accurate predictions in our numerical experiments. However, we can also extend our numerical analysis as an extrapolation problem.
For example, let us consider two trained networks: one network utilizes $AR = 1, 2, 3$ and $7$, while the other network includes $AR = 2, 3, 5$ and $7$ as the training cases together with $\varphi = 10^o, 30^o$ and $60^o$ rounded square cases. 
In both cases, the drag coefficient of ellipse geometries with the aspect ratio outside the training set are predicted reasonably. Figure \ref{fig:ExpolCNN} illustrates the predictions of these cases with an error about  5\%. It is worth noting that the interpolation approach is slightly more accurate than the extrapolation cases. Relevant codes and data for this particular example can be found at GitHub link: \url{https://github.com/TharinduMiyanawala/CNNforCFD}.  
Overall, the above investigation suggests that we can achieve an improved performance by allocating the high-performance computing resources for the FOM cases at the extreme values of the input parameter.

\begin{figure}[h!]
\centering
\begin{subfigure}{\textwidth}
\centering
\includegraphics[trim={138 328 102 180},clip,scale=0.5]{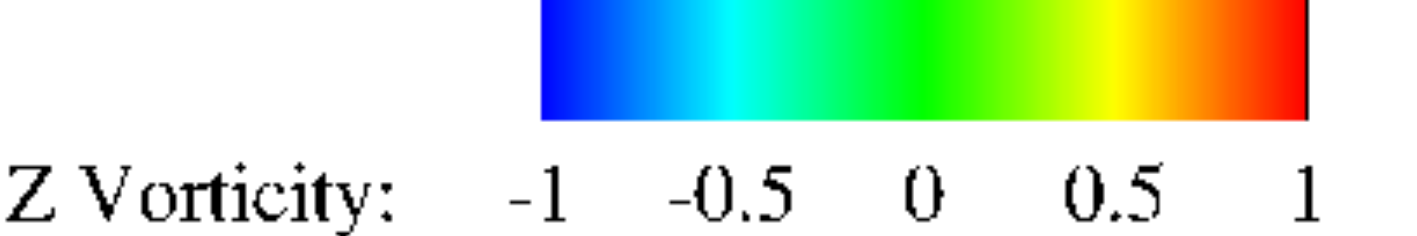}
\end{subfigure} \\ \hspace{5cm}
\begin{subfigure}{0.5\textwidth}
\centering
\includegraphics[trim={30 85 0 255},clip,scale=0.275]{Trapz.eps}
\caption{}
\end{subfigure}~
\begin{subfigure}{0.5\textwidth}
\includegraphics[trim={10 70 0 270},clip,scale=0.275]{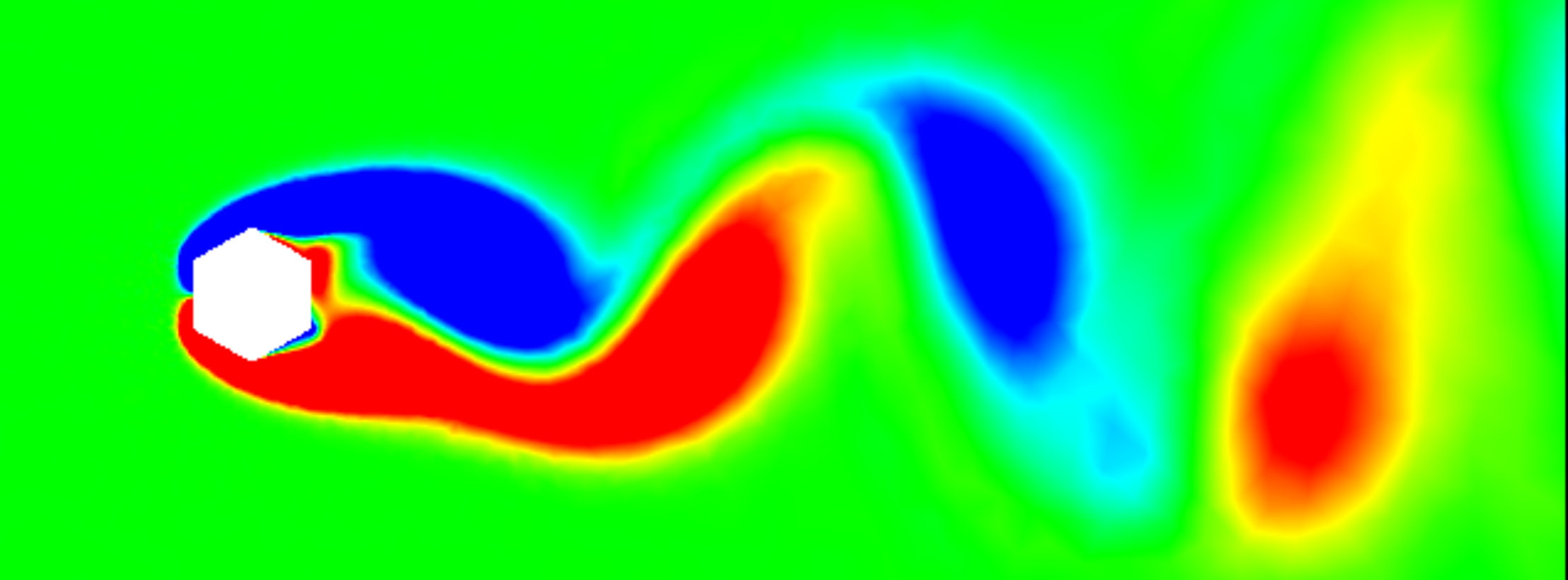}
\caption{}
\end{subfigure}
\begin{subfigure}{0.5\textwidth}
\centering
\includegraphics[trim={10 70 10 270},clip,scale=0.275]{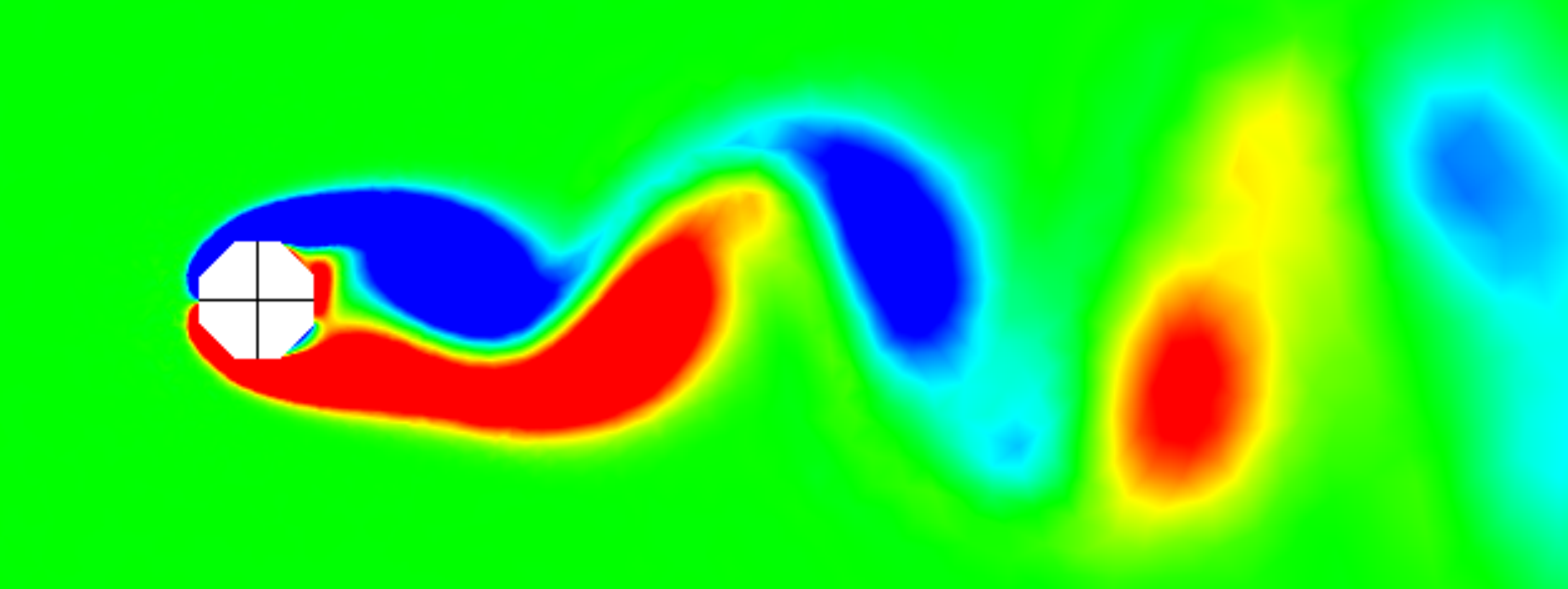}
\caption{}
\end{subfigure}~
\begin{subfigure}{0.5\textwidth}
\includegraphics[trim={0 70 10 270},clip,scale=0.275]{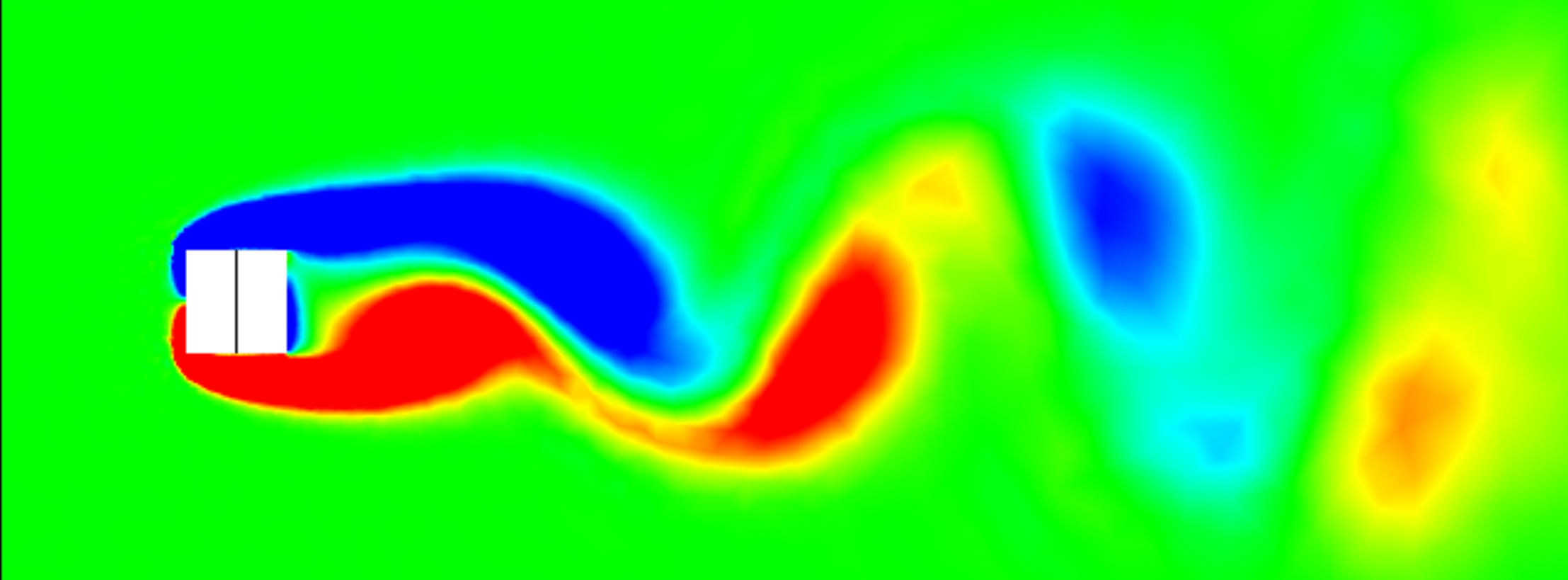}
\caption{}
\end{subfigure}
\begin{subfigure}{0.5\textwidth}
\includegraphics[trim={0 40 10 300},clip,scale=0.275]{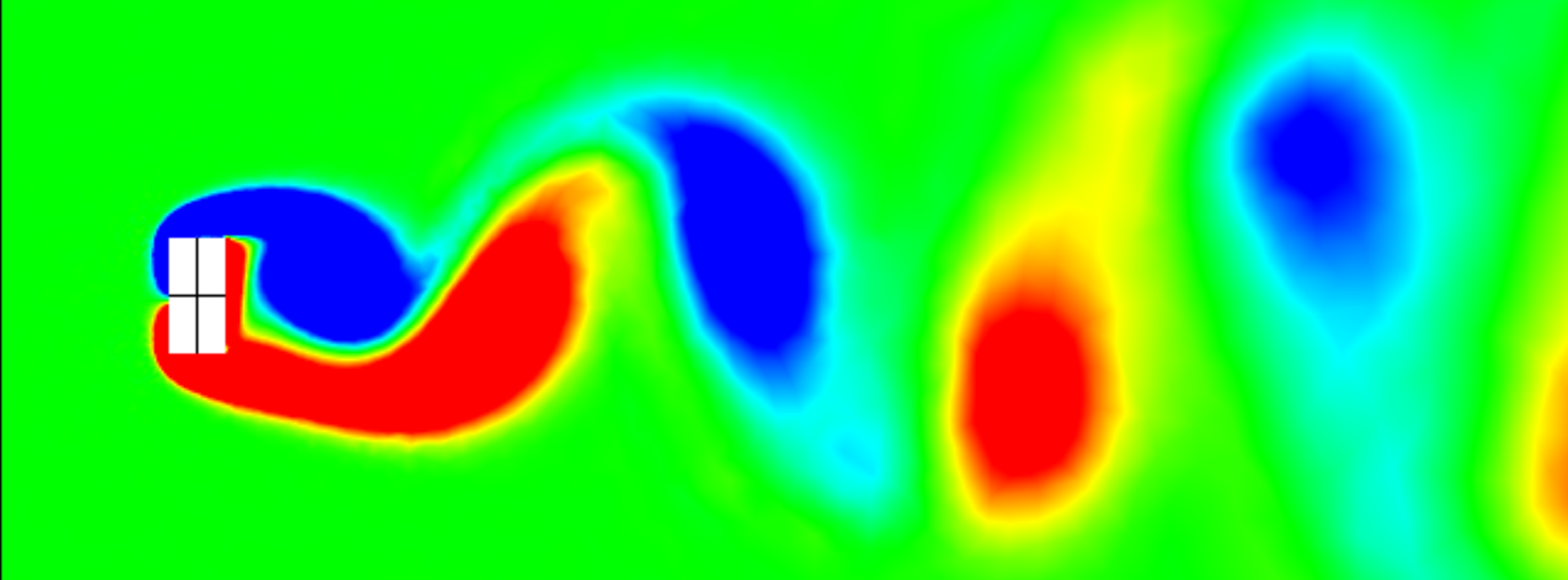}
\caption{}
\end{subfigure}
  \caption{\color{red} Vorticity fields for sharp cornered bluff-bodies:
  (a) trapezium, (b) regular hexagon,
  (c) regular octagon, (d) square and (e) rectangle with aspect ratio $=2$ at $tU_{\infty}/D = 245$. Flow is from left to right. }
  \label{fig:Other_BB}
\end{figure}

Next, we attempt to generalize the force coefficient predictions to other bluff body shapes with sharp corners. In particular, we utilize the same CNN trained presented in Section \ref{sec:Overfitting} to predict the drag coefficient for the bluff body geometries shown in Fig. \ref{fig:Other_BB}, i.e. a symmetric trapezium, a regular hexagon, a regular octagon, a basic square and a rectangular geometry with an aspect ratio of 2. The force predictions and the associated errors are presented in table \ref{Table:Other_BB} in the ascending order of the drag coefficient. The CNN forecasts the drag coefficients with more than 92.5\% accuracy. As expected, the shapes closer to the circle and the rounded square (i.e., sharp square, octagon, and hexagon) have higher accuracy than the trapezium and the rectangle geometries. These results demonstrate a broader prospect of the data-driven flow dynamics predictions via trained deep nets. Using the available full-order or experimental results, one can create a common database for flow dynamic coefficients and use it to train a generalized CNN to predict the flow dynamic parameters of any complex shape for a wide range of flow conditions.

\begin{table}[h!]	 
\centering
\caption{\color{red}Generalization of drag force predictions for sharp-cornered bluff-bodies shown in Fig. \ref{fig:Other_BB}.}
\color{red}
\begin{tabular}{l|r|r|c}
\multicolumn{1}{c|}{\multirow{2}{*}{Bluff-body}} & \multicolumn{2}{c|}{$\overline{C_D}$}                   & \multicolumn{1}{c}{\multirow{2}{*}{Error (\%)}} \\ \cline{2-3}
\multicolumn{1}{c|}{}                            & \multicolumn{1}{c|}{FOM} & \multicolumn{1}{c|}{CNN} & \multicolumn{1}{c}{}   \\ \hline
Trapezium                                        & 1.3584                   & 1.2597                       & 7.20                                         \\
Hexagon                                          & 1.3787                   & 1.3741                       & 0.33                                         \\
Octagon                                          & 1.4196                   & 1.3973                       & 1.57                                         \\
Square                                           & 1.5095                   & 1.4291                       & 5.30                                         \\
Rectangle                                        & 1.8341                   & 1.7186                       & 6.29                                        
\end{tabular}
\label{Table:Other_BB}
\end{table}

Efficient prediction of force coefficients via our data-driven framework has a profound impact in offshore, civil and aeronautical engineering applications. In particular to practical offshore structures, the hydrodynamic coefficients required for the well-known force decomposition model of \cite{lighthill1986fundamentals} can be obtained efficiently from the present CNN-based model which are otherwise expensive and slow via CFD-based simulations or physical experiments. Further extensions and demonstrations are underway for high Reynolds number and practical geometries.

\section{Conclusions}
\label{Conclu}
We have presented a general and efficient data-driven model reduction method based on the deep-learning convolutional 
neural nets and the stochastic gradient descent with momentum term.
We successfully demonstrated the  predictive capability of the proposed deep learning technique  for the force coefficients of a set of bluff body geometries. 
We provided a physical analogy of the stochastic gradient descent method and the iterative correction process of the CNN-based with the simplified form of the Navier-Stokes equation. 
We illustrated the connection of the convolution process in the deep neural networks with the Mori-Zwanzig formalism.
During the prediction process, we feed this CNN-based model with the force coefficients derived for seven bluff bodies using
a full-order Navier-Stokes solver. The CNN is trained to output the force coefficients 
for any perturbed bluff body geometry input. We successfully tuned the neural
network parameters for the best prediction performance with a maximum error $<5\%$.
We found that the predictions are most accurate when we use a single convolution layer with 50 kernels of size $4 \times 4$ followed by an reLU rectifier and one fully-connected layer using 0.01 learning rate for the stochastic gradient descent method.
We assured that the CNN is not overfitted for the training dataset using 
a 1-fold exclusion test and a class-wise training test.
The force coefficient predictions for twenty-one new bluff
bodies are obtained almost in real-time with the relative error threshold of 5\%. The CNN-based prediction enables the design space exploration using a small fraction of time and computational resources compared to full-order simulations.
We have compared the theoretical computational gain for the on-line computation of the proposed CNN-based learning against the FOM analysis based on the finite-element GMRES formulation. The observed performance gain through our numerical experiments is found about $O(10^4)$ for the considered set of cases.
We have shown that the CNN-based learning accurately captures the flow features related to the force predictions with very less input-output data combinations. The CNN-based predictions are accurate for both interpolation and extrapolation from the geometric parameter range of the training set. Finally, we demonstrated that the trained CNN can be used to predict many different geometries.
The efficient prediction of force coefficients using the present data-driven method has a great value for the iterative engineering design and the feedback flow control \cite{yao2017feedback}. 
We hope that this work will help to expand the 
use of deep neural networks in a wider range of CFD applications.

\appendix
\setcounter{equation}{0} 
\setcounter{figure}{0}
\renewcommand{\theequation}{A.\arabic{equation}}
\section*{Appendix A: Transformation of the NS momentum equation to reaction-diffusion equation}
We briefly present the transformation of the NS momentum equation into the reaction-diffusion problem by rewriting the convective and diffusion terms and eliminating the convective operator.
Let $\phi$ be the velocity potential function as $\boldsymbol{u} = \boldsymbol{\nabla}\phi$ and then from the incompressible Navier-Stokes momentum equation (Eq. (\ref{eq:NS})), we get
\begin{align}
\frac{\partial \boldsymbol{\nabla}\phi}{\partial t} + (\boldsymbol{\nabla}\phi \cdot \boldsymbol{\nabla}) \boldsymbol{\nabla}\phi = -\frac{1}{\rho}\boldsymbol{\nabla}p + \nu \boldsymbol{\nabla^2}\boldsymbol{\nabla}\phi. \label{eq:VelPot}
\end{align}
Using the vector identity: 
$\boldsymbol{\nabla}(\boldsymbol{a}.\boldsymbol{a}) =2\left(\boldsymbol{a}\times(\boldsymbol{\nabla}\times\boldsymbol{a})+(\boldsymbol{a}.\boldsymbol{\nabla})\boldsymbol{a}\right)$ and $\boldsymbol{\nabla} \times \boldsymbol{\nabla}\phi = 0$,  we get $(\boldsymbol{\nabla}\phi \cdot \boldsymbol{\nabla}) \boldsymbol{\nabla}\phi = \frac{1}{2}\boldsymbol{\nabla}(\boldsymbol{\nabla}\phi \cdot \boldsymbol{\nabla}\phi)$ and then substituting into Eq. (\ref{eq:VelPot}), one can obtain
\begin{align}
\boldsymbol{\nabla}\left(\frac{\partial\phi}{\partial t} +\frac{\boldsymbol{\nabla}\phi \cdot \boldsymbol{\nabla}\phi}{2}\right)=\boldsymbol{\nabla}\left(-\frac{p}{\rho}+\nu{\nabla^2}\phi\right),
\end{align}
which can be further simplified as
\begin{align}
\frac{\partial\phi}{\partial t} +\frac{\boldsymbol{\nabla}\phi \cdot \boldsymbol{\nabla}\phi}{2}=-\frac{\Delta p}{\rho}+\nu{\nabla^2}\phi,
\label{eq:potenNS}
\end{align}
where $\Delta p$ is the pressure difference with respect to some reference pressure.
By introducing the variable transformation $\phi=-2\nu \ln{\psi}$ in Eq.(\ref{eq:potenNS}), we obtain:
\begin{align}
\frac{1}{\psi}\frac{\partial\psi}{\partial t} -\frac{\nu}{\psi^2}\boldsymbol{\nabla}\psi \cdot \boldsymbol{\nabla}\psi=\frac{\Delta p}{2\mu}+\nu{\nabla^2}(\ln{\psi}).
\end{align}
By substituting the relation ${\nabla^2}(\ln{\psi})=\left(-\frac{1}{\psi^2}(\boldsymbol{\nabla}\psi \cdot \boldsymbol{\nabla}\psi)+\frac{1}{\psi}{\nabla^2}\psi\right)$, we construct the reaction-diffusion equation in terms of scalar function $\psi$ of a moving fluid element from the Navier-Stokes momentum equation:
\begin{align}
\frac{\partial\psi}{\partial t} - \nu{\nabla^2}\psi = \frac{\Delta p}{2\mu}\psi.
\label{eq:Reac-diff}
\end{align}
The above hyperbolic differential form is often called the Liouville equation, whose 
characteristics are integral curves of the nonlinear ordinary differential equation.
In Appendix B, we present an integral solution to the above 
reaction-diffusion equation for constructing the moment map.
\appendix
\setcounter{equation}{0} 
\setcounter{figure}{0}
\renewcommand{\theequation}{B.\arabic{equation}}
\section*{Appendix B: An explicit integral solution of the reaction-diffusion equation}
We can construct a general integral solution for the reaction-diffusion system Eq. (\ref{eq:Reac-diff}) via Green's function.
Let $\mathcal{L}$ be the operator $\mathcal{L} = \frac{\partial}{\partial t}-\nu\boldsymbol{\nabla^2}$. The differential equations for $\psi(\xx,t)$ and the Green's function, $G(\xx,t;\boldsymbol{\xi},\tau)$ for $\xx,\boldsymbol{\xi}\in \mathcal{D}$ and $t,\tau \geq 0$ can be expressed as:
\begin{align}
\mathcal{L}\psi(\xx,t) = \frac{\Delta p}{2\mu}\psi(\xx,t), \label{eq:Reac-diff-opt} \\
\mathcal{L}G(\xx,t;\boldsymbol{\xi},\tau)=\delta (\xx-\boldsymbol{\xi})\delta(t-\tau), \label{eq:Reac-diff-Green} 
\end{align}
where $\mathcal{D}$ is the fluid domain of interest and $\boldsymbol{\xi}$ and $\tau$ are the dummy variables.
Multiplying Eq.~(\ref{eq:Reac-diff-opt}) by $G$ and  by Eq.~(\ref{eq:Reac-diff-Green}) $\psi$ and subtracting Eq.~(\ref{eq:Reac-diff-Green}) from Eq.~(\ref{eq:Reac-diff-opt}),  we get
\begin{align}
G\mathcal{L}\psi - \psi\mathcal{L}G = G\frac{\Delta p}{2\mu}\psi - \delta (\xx-\boldsymbol{\xi})\delta(t-\tau)\psi.
\end{align}
Integrating with respect to $\xx$ and $t$, the space-time integral form is constructed as follows:
\begin{align}
\int_0^{\infty}\int_V [G\mathcal{L}\psi - \psi \mathcal{L}G] \, dV dt = \int_0^{\infty}\int_V G\frac{\Delta p}{2\mu}\psi \, dV dt - \psi(\boldsymbol{\xi},\tau).
\label{eq:InteGreen}
\end{align}
The left-hand side can be further simplified using the Green's second identity
\begin{equation}
\begin{split}
\int_0^{\infty}\int_V [G\mathcal{L}\psi - \psi\mathcal{L}G]\,dV dt = \int_V \int_0^{\infty}\left[G\frac{\partial\psi}{\partial t} - \psi \frac{\partial G}{\partial t}\right]\,dtdV \\ -\nu \int_0^{\infty}\int_V \left[G{\nabla^2}\psi  - \psi {\nabla^2}G \right]\,dV dt \\
= -\int_V G\psi(\xx,0)\,dV-2\int_0^{\infty}\int_V \psi \frac{\partial G}{\partial t}\,dV dt -\nu \int_0^{\infty} \oint_S \left[G\boldsymbol{\nabla}\psi  - \psi \boldsymbol{\nabla}G \right]\boldsymbol{n}\,dSdt.
\end{split}
\label{eq:LHSGreen}
\end{equation}
By equating Eq.~(\ref{eq:InteGreen}) and Eq.~(\ref{eq:LHSGreen}), for $\psi(\boldsymbol{\xi},\tau)$ we obtain the explicit expression for the transformed scalar variable $\psi$:
\begin{equation}
\begin{split}
\psi(\boldsymbol{\xi},\tau)= \int_0^{\infty}\int_V G\frac{\Delta p}{2\mu}\psi\, dV dt +\int_V G\psi(\xx,0)\,dV + 2\int_0^{\infty}\int_V \psi \frac{\partial G}{\partial t}\,dV dt \\ +\nu \int_0^{\infty} \oint_S \left[G\boldsymbol{\nabla}\psi  +\psi \boldsymbol{\nabla}G \right]\boldsymbol{n}\,dSdt.
\end{split}
\end{equation}
By exchanging $(\boldsymbol{\xi},\tau)$ with $(\xx,t)$ and setting the 3rd term of right-hand side to zero, we get the general solution of non-homogeneous diffusion equation \cite{haberman1983elementary}:
\begin{equation}
\begin{split}
\psi(\xx,t)= \int_0^{\infty}\int_V G\frac{\Delta p}{2\mu}\psi(\xx,t)\, dV_{\boldsymbol{\xi}}d\tau +\int_V G\psi(\xx,0)\,dV_{\boldsymbol{\xi}} \\ +\nu \int_0^{\infty} \oint_S \left[G\boldsymbol{\nabla}_{\boldsymbol{\xi}} \psi  +\psi \boldsymbol{\nabla}_{\boldsymbol{\xi}} G \right]\boldsymbol{n}\,dSd\tau.
\label{eq:nonhomog}
\end{split}
\end{equation}
By observing the above equation, 
one can construct an input-output mapping for the sets of scalar functions and the Green's function.
Deep neural networks can help to construct these functions hierarchically, in which higher-level feature functions
can be formed as a multilayer stack of simple lower-level ones.
\appendix
\setcounter{equation}{0} 
\setcounter{figure}{0}
\renewcommand{\theequation}{C.\arabic{equation}}
\renewcommand{\thefigure}{C.\arabic{figure}}
\section*{Appendix C: The Mori-Zwanzig projection operators}
By considering the operator $L=\frac{\Delta p}{2\mu}+\nu\nabla^2$, the reaction-diffusion Eq. (\ref{eq:Reac-diff}) derived from the Navier-Stokes momentum equation can be written as Liouville form:
\begin{equation}
\frac{\partial{\psi}}{\partial t}=L{\psi},
\label{eq:Liouvillean}
\end{equation}
whose characteristics satisfy the $N$-dimensional 
semi-discrete nonlinear ordinary differential equation. 
Let $\psi(\chi_0,t) = g(\chi)$, where $\chi \in\mathbb{R}^N$ denote the modes of $\psi$ comprising the slow (retained) degrees of freedom $\widehat{\chi}$ and 
the fast (eliminated) degrees of freedom $\widetilde{\chi}$.
We denote the component of $\psi$ corresponding to the resolved mode by $\widehat{\psi}(t) = \psi(\widehat{\chi}_0,t) = g (\widehat{\chi})$. 
The solution of Eq. (\ref{eq:Liouvillean}) for the resolved mode is given by $\widehat{\psi}=e^{Lt}\widehat{\psi}(0)$, whereas $e^{Lt}$ is the evolution operator. 
The standard semi-group notation is used to construct the flow map i.e., $\widehat{\psi}(\widehat{\chi}_0,t)=e^{Lt} g(\widehat{\chi}_0)$.
For brevity, we denote $\widehat{\psi}(0)$ by $\widehat{\psi}_0$ hereafter. 
Let $P$ be a projection of $\psi$ in the direction of $\widehat{\psi}$ 
such that
\begin{equation}
P\psi \equiv \frac{(\widehat{\psi}_0,{\psi}_0)}{(\widehat{\psi}_0,\widehat{\psi}_0)}\widehat{\psi},
\label{eq:Projection}
\end{equation}
and also let $Q=I-P$ be another projector onto an orthogonal subspace. The projection operators $P$ and $Q$ follow the self-adjoint condition and satisfy the orthogonality condition $QP=0$.
While $P$ projects the particular variable onto the slow variables of the system, $Q$ projects the variable onto the fast variables of the dynamical system.
Using Dyson's formula \cite{morriss2013statistical}:
\begin{equation}
e^{Lt} = e^{QLt}+\int_0^t e^{L(t-t_1)}PLe^{QLt_1}dt_1.
\label{eq:Dyson}
\end{equation}
and substituting it to Eq. (\ref{eq:Liouvillean}), we obtain:
\begin{equation}
\frac{\partial \widehat{\psi}}{\partial t}=e^{Lt}L\widehat{\psi}_0 = e^{Lt}(Q+P)L\widehat{\psi}_0 =e^{Lt}PL\widehat{\psi}_0+e^{Lt}QL\widehat{\psi}_0.
\label{eq:MZIntermediate}
\end{equation}
Using Eq. (\ref{eq:Projection}) and Eq. (\ref{eq:Dyson}), 
the projection terms are:
\begin{align}
e^{Lt}PL\widehat{\psi}_0 = \frac{(L\widehat{\psi}_0,{\psi}_0)}{(\widehat{\psi}_0,\widehat{\psi}_0)}e^{Lt}\widehat{\psi}_0 = \Theta\widehat{\psi}, \\
e^{Lt}QL\widehat{\psi}_0 = e^{QLt}QL\widehat{\psi}_0 + \int_0^t e^{L(t-t_1)}PLe^{QLt_1}QL\widehat{\psi}_0\,dt_1.
\end{align}
By defining $F(\widehat{\chi}_0, t)\equiv e^{QLt}QL\widehat{\psi}_0$ and $PLe^{QLt}QL\widehat{\psi}_0=-K(t)\widehat{\psi}_0$,  we can write Eq. (\ref{eq:MZIntermediate}) as
\begin{equation}
\frac{\partial \widehat{\psi}}{\partial t}=\underbrace{\Theta\widehat{\psi}}_\text{Markovian} + \underbrace{\int_0^t K(t_1)\widehat{\psi}(t-t_1)\,dt_1}_\text{Convolution (memory)} + \underbrace{F(\widehat{\chi}_0,t)}_\text{Fluctuation}
\label{eq:MZFormalism}
\end{equation}
%
%
which is referred to as the Mori-Zwanzig equation for the coarse-graining of a general dynamical system \cite{mori1965transport,zwanzig1973nonlinear,givon2004extracting}. 
While the first term in the right-hand side is a function of the instantaneous value of $\psi$, the second term depends on the time history of $\psi$ in the range $[0,t]$ via convolution (memory) effect. The third term represents the effect of fluctuations and by construction, it satisfies the orthogonal dynamics equation, $\frac{\partial F(\widehat{\chi}_0,t)}{\partial t} = QLF(\widehat{\chi}_0,t)$. 
This formalism provides a theoretical framework to make systematic approximations of memory kernel effects during the CNN-based learning.
\section*{Acknowledgements}
The first author thanks the National Research Foundation and Keppel Corporation, Singapore for supporting the work done in the Keppel-NUS Corporate Laboratory.

\section*{References}

\bibliography{CNNRef}

\end{document}